\documentclass[a4paper,11pt]{article}
\pdfoutput=1
\usepackage{jheppub}
\usepackage[utf8]{inputenc}

\usepackage{pdflscape}
\usepackage{tikz}
\usepackage{tikz-3dplot}
\usetikzlibrary{positioning}
\usetikzlibrary{arrows}
\usetikzlibrary{decorations.pathreplacing}
\usetikzlibrary{shapes.geometric}
\usetikzlibrary{calc}
\usepackage{todonotes}
\usepackage{enumitem}
\usepackage{afterpage}
\usepackage{amsmath}
\usepackage{tensor}

\usepackage{mathtools}

\usepackage{MnSymbol}
\usepackage{oubraces}
\usepackage{tikz}
\usetikzlibrary{shapes.geometric}
\tikzset{gauge1/.style={draw=none,minimum size=0.4cm,fill=white,circle, draw}}
\tikzset{gauge5/.style={draw=none,minimum size=0.6cm,fill=white,circle, draw}}
\tikzset{supergauge/.style={draw=none,minimum size=0.9cm,fill=white,circle, draw}}
\tikzset{bluenode/.style={draw=none,minimum size=0.4cm,fill=blue,circle, draw}}
\tikzset{rednode/.style={draw=none,minimum size=0.4cm,fill=red,circle, draw}}
\tikzset{gauge3/.style={draw=none,minimum size=0.4cm,fill=white,circle, draw}}
\tikzset{dotsize/.style={draw=none,minimum size=0.6pt,fill=black,circle,inner sep=1pt, draw}}
\tikzset{mini/.style={draw=none,minimum size=1pt,fill=white,circle,inner sep=3pt, draw}}
\tikzset{miniG/.style={draw=none,minimum size=1pt,fill=black,circle,inner sep=3pt, draw}}
\tikzset{cyane/.style={draw=none,minimum size=0.4cm,fill=cyan,circle, draw}}
\tikzset{pinklinet/.style={draw=none,minimum size=0.4cm,fill=magenta,circle, draw}}
\tikzset{greenlinet/.style={draw=none,minimum size=0.4cm,fill=green,circle, draw}}
\tikzset{blacknode/.style={draw=none,minimum size=0.4cm,fill=black,circle, draw}}
\tikzset{brownlinet/.style={draw=none,minimum size=0.4cm,fill=olive,circle, draw}}
\tikzset{magicmintlinet/.style={draw=none,minimum size=0.4cm,fill=red,circle, draw}}
\tikzset{orangeet/.style={draw=none,minimum size=0.4cm,fill=orange,circle, draw}}
\tikzset{grayet/.style={draw=none,minimum size=0.4cm,fill=gray,circle, draw}}
\tikzset{blueet/.style={draw=none,minimum size=0.4cm,fill=blue,circle, draw}}
\tikzset{flavour2/.style={draw=none,minimum size=0.8cm,fill=white, regular polygon,regular polygon sides=4,draw}}
\tikzset{flavour2/.style={draw=none,minimum size=0.6cm,fill=white, regular polygon,regular polygon sides=4,draw}}
\tikzset{redsquare/.style={draw=none,minimum size=0.6cm,fill=red, regular polygon,regular polygon sides=4,draw}}
\tikzset{bluesquare/.style={draw=none,minimum size=0.6cm,fill=blue, regular polygon,regular polygon sides=4,draw}}
\tikzset{greenflavor/.style={draw=none,minimum size=0.6cm,fill=green, regular polygon,regular polygon sides=4,draw}}
\tikzset{brownflavor/.style={draw=none,minimum size=0.6cm,fill=brown, regular polygon,regular polygon sides=4,draw}}
\tikzset{pinkflavor/.style={draw=none,minimum size=0.6cm,fill=magenta, regular polygon,regular polygon sides=4,draw}}
\tikzset{grayflavor/.style={draw=none,minimum size=0.6cm,fill=gray, regular polygon,regular polygon sides=4,draw}}
\tikzset{none/.style={draw=none}}
\tikzset{new edge style 1/.style={dashed}}
\tikzset{dashedline/.style={dashed}}
\tikzset{brace1/.style={decorate,decoration={brace,amplitude=5pt,mirror}}}
\tikzset{bluee/.style={line width=0.5mm,blue}}
\tikzset{orangee/.style={line width=0.5mm,orange}}
\tikzset{magentae/.style={line width=0.5mm,magenta}}
\tikzset{rede/.style={line width=0.5mm,red}}
\tikzset{greene/.style={line width=0.5mm,green}}
\tikzset{darke/.style={line width=0.5mm,black}}
\tikzset{cyaneX/.style={line width=0.5mm,cyan}}
\tikzset{new edge style 3/.style={dashed,red}}
\tikzset{magicmintline/.style={line width=0.5mm,gray}}
\tikzset{brownline/.style={line width=0.5mm,brown}}
\tikzset{greenline/.style={line width=0.5mm,green}}
\tikzset{oliveline/.style={line width=0.5mm,green}}
\tikzset{darkgreenline/.style={line width=0.5mm,olive}}
\tikzset{pinkline/.style={line width=0.5mm,magenta}}
\tikzset{dottedz/.style={line width=0.5mm,black,dotted}}
\tikzset{pinkline2/.style={line width=0.5mm,magenta,dotted}}
\tikzset{brace2/.style={decorate,decoration={brace,amplitude=5pt}}}
\tikzset{reddotted/.style={line width=0.5mm,red,dotted}}
\tikzset{bluedotted/.style={line width=0.5mm,blue,dotted}}
\tikzset{magicmintdotted/.style={line width=0.5mm,gray,dotted}}
\tikzset{greendotted/.style={line width=0.5mm,green,dotted}}
\tikzset{browndotted/.style={line width=0.5mm,brown,dotted}}
\tikzset{arrowed/.style={line width=0.5mm, ->}}
\pgfdeclarelayer{edgelayer}
\pgfdeclarelayer{nodelayer}
\usetikzlibrary{decorations.pathreplacing}
\pgfsetlayers{edgelayer,nodelayer,main}
\tikzset{hasse/.style={circle, fill,inner sep=2pt}}
\tikzset{hasseb/.style={circle, fill=black!40,inner sep=4pt}}
\tikzset{hassebr/.style={circle, fill=red!40,inner sep=4pt}}
\tikzset{hassegray/.style={circle, fill=black!10,inner sep=2pt}}
\tikzset{linegray/.style={black!10}}
\tikzset{gauge/.style={inner sep=1mm,draw=none,fill=white,minimum size=2mm,circle, draw}}
\tikzset{gauger/.style={inner sep=1mm,draw=none,fill=goodred,minimum size=2mm,circle, draw}}
\tikzset{gaugeo/.style={inner sep=1mm,draw=none,fill=goodorange,minimum size=2mm,circle, draw}}
\tikzset{gaugeb/.style={inner sep=1mm,draw=none,fill=blue,minimum size=2mm,circle, draw}}
\tikzset{gaugeg/.style={inner sep=1mm,draw=none,fill=goodgreen,minimum size=2mm,circle, draw}}
\tikzset{gaugec/.style={inner sep=1mm,draw=none,fill=cyan,minimum size=2mm,circle, draw}}
\tikzset{gaugem/.style={inner sep=1mm,draw=none,fill=goodmagenta,minimum size=2mm,circle, draw}}
\tikzset{flavour/.style={draw=none,minimum size=0.3mm,fill=white, regular polygon,regular polygon sides=4,draw}}
\tikzset{flavourr/.style={draw=none,minimum size=0.3mm,fill=red, regular polygon,regular polygon sides=4,draw}}
\tikzset{sev/.style={inner sep=1mm,draw=none,fill=white,minimum size=4mm,circle, draw}}
\tikzset{sevR/.style={inner sep=1mm,draw=none,fill=white,minimum size=4mm,circle, draw, red,fill=white}}
\tikzset{sevB/.style={inner sep=1mm,draw=none,fill=white,minimum size=4mm,circle, draw, blue,fill=white}}
\tikzset{sevG/.style={inner sep=1mm,draw=none,fill=white,minimum size=4mm,circle, draw, goodgreen,fill=white}}
\tikzset{NS/.style={circle, fill=red,inner sep=3pt}}
\tikzset{rec/.style={draw, rectangle, minimum size=2mm,align=center}}
\tikzset{quadruple/.style={double distance=#1-\pgflinewidth,thick,
        postaction={draw,thick,double distance=#1/3-\pgflinewidth}},
    quadruple/.default=.5em}
\tikzset{doublea/.style={double distance=1/4*(#1-\pgflinewidth),thick},
    doublea/.default=.5em}
\colorlet{c1}{white}
\colorlet{c3}{orange!10}
\colorlet{c6}{red!20}

\newcommand{\Lie}[1]{\operatorname{\textsl{#1}}}
\newcommand{\lie}[1]{\operatorname{\mathfrak{#1}}}
\newcommand{\GL}{\Lie{GL}}

\newcommand{\su}{\lie{su}}
\newcommand{\un}{\lie u}

\newcommand{\n}{\lie n}
\newcommand{\p}{\lie p}

\newcommand{\SL}{\Lie{SL}}
\newcommand{\SU}{\Lie{SU}}
\newcommand{\Un}{\Lie{U}}

\newcommand{\uri}{\tensor{\operatorname{Q}}{_{G}}}
\newcommand{\uli}{\tensor[_G]{\operatorname{Q}}{}}
\newcommand{\urip}{\tensor{\operatorname{Q}}{_{G,P}}}
\newcommand{\ulip}{\tensor[_{G,P}]{\operatorname{Q}}{}}
\newcommand{\ucon}{\tensor{\operatorname{R}}{_G}}
\newcommand{\uconp}{\tensor{\operatorname{R}}{_{G,P}}}

\definecolor{goodgreen}{RGB}{55,169,49}
\definecolor{goodred}{RGB}{230,0,0}
\definecolor{goodorange}{RGB}{255,200,0}
\definecolor{goodmagenta}{RGB}{255,150,255}
\definecolor{goodyellow}{RGB}{160,160,0}

\newcommand{\convexpath}[2]{
  [   
  create hullcoords/.code={
    \global\edef\namelist{#1}
    \foreach [count=\counter] \nodename in \namelist {
      \global\edef\numberofnodes{\counter}
      \coordinate (hullcoord\counter) at (\nodename);
    }
    \coordinate (hullcoord0) at (hullcoord\numberofnodes);
    \pgfmathtruncatemacro\lastnumber{\numberofnodes+1}
    \coordinate (hullcoord\lastnumber) at (hullcoord1);
  },
  create hullcoords
  ]
  ($(hullcoord1)!#2!-90:(hullcoord0)$)
  \foreach [
  evaluate=\currentnode as \previousnode using \currentnode-1,
  evaluate=\currentnode as \nextnode using \currentnode+1
  ] \currentnode in {1,...,\numberofnodes} {
    let \p1 = ($(hullcoord\currentnode) - (hullcoord\previousnode)$),
    \n1 = {atan2(\y1,\x1) + 90},
    \p2 = ($(hullcoord\nextnode) - (hullcoord\currentnode)$),
    \n2 = {atan2(\y2,\x2) + 90},
    \n{delta} = {Mod(\n2-\n1,360) - 360}
    in 
    {arc [start angle=\n1, delta angle=\n{delta}, radius=#2]}
    -- ($(hullcoord\nextnode)!#2!-90:(hullcoord\currentnode)$) 
  }
}

\newtheorem{remark}{Remark}

\usepackage[makeroom]{cancel}

\preprint{}

\title{Complex Symplectic Contractions and 3d Mirrors}

\author[a,c]{Andrew Dancer,}
\author[a]{Julius F. Grimminger,}
\author[b]{Johan Martens,}
\author[a]{and Zhenghao Zhong}
\affiliation[a]{Mathematical Institute, University of Oxford,\\
Andrew Wiles Building, Woodstock Road, Oxford, OX2 6GG, UK}
\affiliation[b]{School of Mathematics and Maxwell Institute, University of Edinburgh, James Clerk Maxwell Building, Edinburgh EH9 3FD,
Scotland, UK}
\affiliation[c]{Jesus College, Oxford OX1 3DW, UK}
\emailAdd{dancer@maths.ox.ac.uk}
\emailAdd{julius.grimminger@maths.ox.ac.uk}
\emailAdd{johan.martens@ed.ac.uk}
\emailAdd{zhenghao.zhong@maths.ox.ac.uk}

\abstract{
We propose magnetic quivers for the complex-symplectic contraction spaces, which
are related to implosions and have a natural interpretation in terms of the Moore-Tachikawa category. We use 3-d mirrors to
provide computational checks.}

\begin{document}

\maketitle

\section{Introduction and Summary}
Complex-symplectic varieties, and in particular symplectic singularities \cite{beauville2000symplectic}, have been at the heart of geometric representation theory for many years. In physics these spaces show up prominently as Higgs branches \cite{Hitchin:1986ea} of supersymmetric quantum field theories (SQFTs) with 8 supercharges in various dimensions (e.g.\ 3d $\mathcal{N}=4$, $4d$ $\mathcal{N}=2$, $5d$ $\mathcal{N}=1$, or $6d$ $\mathcal{N}=(1,0)$ SQFTs), or as Coulomb branches of 3d $\mathcal{N}=4$ SQFTs \cite{Cremonesi:2013lqa,Bullimore:2015lsa,Nakajima:2015txa,Braverman:2016wma, MR4020310}.

The construction of such a space as the 3d $\mathcal{N}=4$ Coulomb branch of a \emph{quiver gauge theory} is extremely useful, as it allows us to compute a range of its properties via combinatorial operations on its quiver. From now on we will refer to such a quiver $\mathsf{Q}$, whose 3d $\mathcal{N}=4$ Coulomb branch $\mathcal C(\mathsf{Q})$ is a variety $X$, as a \emph{magnetic quiver} for $X$. We use $\mathcal{H}(\mathsf{Q})$ to denote the Higgs branch of the quiver, which is the symplectic dual \cite{Webster1407} of $\mathcal C(\mathsf{Q})$.

On the other hand, we may also be able to construct our space in question directly as a Higgs branch of a quiver, which we then refer to as an electric quiver for our space. Two quivers $\mathsf{Q}$ and $\mathsf{Q}^\vee$ are called a $3d$ mirror pair, if among other things (see \cite{Intriligator:1996ex}) the Higgs branch of one is the Coulomb branch of the other and vice versa.

In previous work \cite{Dancer:2020wll,Bourget:2021zyc,Bourget:2021qpx} we investigated the magnetic quivers of complex-symplectic implosion spaces \cite{dancer2013implosion}.
We remark here that in the $A_n$ case such spaces have a stratified hyper-K\"ahler structure
and this is expected to be true in general, hence they are often referred to as hyper-K\"ahler implosions.

In the current paper we extend this to the related construction of complex-symplectic contraction \cite{2024arXiv240107920D}. The contraction of a complex-symplectic space has the same dimension as the original space but the symmetry is enhanced by an abelian group (in the simplest case by a maximal torus factor). As for implosion, the complex-symplectic contraction operation is the complex-symplectic analogue of operations in equivariant algebraic \cite{Popov:1986} and real symplectic geometry \cite{HMM:2017}.  Contractions may be understood in terms of a universal example, which is a 
complex-symplectic quotient of the square of the universal implosion. In the Moore-Tachikawa category of complex-symplectic varieties \cite{Moore:2011ee} the contraction morphism is the composition of two implosion morphisms. A complex-symplectic quotient by abelian factors which are diagonal in the symmetry of two spaces (which is the case in this construction) leaves behind an abelian factor. Hence the global symmetry of contraction spaces are larger than `usual'.

We construct the magnetic quivers of contraction spaces as an operation on the implosion magnetic quivers, realising the complex-symplectic quotient on the Coulomb branch. We give background on these quotients in Appendix \ref{app:HKQ}. We note that these
are complex-symplectic quotients by a complex reductive group, so can be interpreted as hyper-K\"ahler quotients.

To conclude the introduction we mention some 
advantages of having a magnetic quiver description $\mathsf{Q}$ of a variety $X$:

\begin{enumerate}
   \item The stratification into symplectic leaves \cite{kraft1980minimal,kaledin2006symplectic} (and their transverse slices) can be obtained from a quiver subtraction algorithm \cite{Cabrera:2018ann,Bourget:2019aer,Bourget:2022ehw,Bourget:2022tmw} and more recently from a `decay and fission' algorithm \cite{Bourget:2023dkj,Bourget:2024mgn}.
    \item Computing the Higgs branch of the magnetic quiver $\mathsf{Q}$ (which is well defined if the quiver is simply-laced) provides the symplectic dual $X^\vee$ of the variety $X$.
    \item If the quiver $\mathsf{Q}$ has a symmetry under exchanging nodes/legs, this implies that the variety $X$ has a discrete symmetry. Quotienting this discrete symmetry can be achieved as a discrete operation on $\mathsf{Q}$ \cite{Hanany:2018dvd,Hanany:2018cgo,Bourget:2020bxh,Hanany:2023uzn}.
    \item Deformations (and the Namikawa Weyl group) of the symplectic dual $X^\vee$ can be studied using a subtraction algorithm on the magnetic quiver \cite{Bourget:2023uhe}.
    \item By construction, if $X=\mathcal{C}(\mathsf{Q})$, $X$ has only symplectic singularities \cite{2020arXiv200501702W}.
\end{enumerate}

\subsection*{Plan of the paper}

The paper is organized as follows. In Section \ref{sec:1} we review the Moore-Tachikawa category and how the implosion and contraction constructions fit into this framework. We provide magnetic quivers of contraction spaces and explore the symplectic duals of the contraction spaces. In Section \ref{sec:2} we discuss 3d mirror pairs, using the quiver description of the $\SU(n)$ implosion, and 
provide computational confirmation of our proposals for magnetic quivers. We also provide some physical implications of how gauging diagonal subgroups can expand the landscape of 3d mirror pairs. In Section \ref{sec:3} we discuss D-type contractions. In the Appendices we provide background material.

\section{Magnetic Quivers for Contraction Spaces}
\label{sec:1}

In this section we first review some basics of the Moore-Tachikawa category and how implosion and contraction spaces show up as morphisms. We then construct explicit magnetic quivers for contraction spaces, and provide a variety of checks.

\subsection{The Moore-Tachikawa Category}
In this section we study certain morphisms in the Moore-Tachikawa category \cite{Moore:2011ee},\footnote{The objects in the category which Moore and Tachikawa call \emph{HS} are semi-simple Lie groups. Here we allow for any reductive group, in particular abelian factors, but still refer to it as Moore-Tachikawa category.} whose objects are reductive groups, and morphisms are complex-symplectic varieties whose global symmetry includes the source and target:
\begin{equation}
    \begin{tikzpicture}
        \node (1) at (0,0) {$G_1$};
        \node (2) at (2,0) {$G_2$.};
        \draw[thick,->] (1)--(2);
        \node at (1,0.3) {$X_{12}$};
    \end{tikzpicture}\;
\end{equation}
Composition of morphisms is given via a complex-symplectic quotient:
\begin{equation}
    \begin{tikzpicture}
        \node (1) at (0,0) {$G_1$};
        \node (2) at (2,0) {$G_2$};
        \node (3) at (4,0) {$G_3$,};
        \draw[thick,->] (1)--(2);
        \draw[thick,->] (2)--(3);
        \node at (1,0.3) {$X_{12}$};
        \node at (3,0.3) {$X_{23}$};
        \draw[thick,->] (1) .. controls (2,-1) .. (3);
        \node at (2,-1.1) {$X_{13}=X_{23}\circ X_{12}$};
    \end{tikzpicture}\;
\end{equation}
with
\begin{equation} \label{composition}
    X_{13}=X_{23}\circ X_{12}=(X_{12}\times X_{23})/\!/\!/ \Delta G_2\;,
\end{equation}
where $\Delta G_2$ denotes the diagonally embedded $G_2$ in the symmetry group $G_1 \times G_2 \times G_2 \times G_3$
of $X_{12} \times X_{23}$.

In general, this quotient will have a residual action of $G_1 \times G_3$, and hence be a morphism from $G_1$ to $G_3$ as required. 
However, if we allow the groups $G$ to have nontrivial centers, then we may acquire more symmetries.
In particular, if the middle group $G_2$ is Abelian then $X_{13}$ actually has an action of $G_1 \times G_2 \times G_3$
because the anti-diagonally embedded $G_2$ commutes with the diagonally embedded $G_2$ and hence descends to the quotient.

\begin{remark}
We note that there is a question in general as to how to interpret the quotient (\ref{composition}). We shall view this as the
Geometric Invariant Theory quotient by $\Delta G_2$ of the zero level set for the complex-symplectic moment map for this group.
In cases of physical interest, the varieties are expected to be Coulomb branches and so affine varieties, in which case the quotient will exist as an affine variety since we are taking our groups to be reductive.
\end{remark}

\begin{remark}
It may be possible to construct higher morphisms in the Moore-Tachikawa picture, with the role of 2-morphisms
between morphisms $X$ and $Y$ in $\operatorname{Hom}(G_1,G_2)$ being played by Lagrangian correspondences $L \subset X^{-} \times Y$ 
with a $G_1 \times G_2$ action where $X^{-}$
refers to $X$ equipped with the opposite symplectic structure. Floer homology groups $HF(L_1,L_2)$ might
be used to define higher morphisms. We refer the reader to Cazassus \cite{Cazassus:2023} for a further discussion of this topic.
\end{remark}

\subsection{Implosion and Contraction Spaces}

\paragraph{Universal Implosion} Let $G$ be a simple group. The universal (right) implosion is a space $Q_G$ of (complex) dimension $\dim G +{\rm rank} \; G$ with an action of
$G \times T$, where $T$ is the maximal complex torus in $G$. Complex-symplectic reduction by $T$ at level zero gives the nilpotent cone, with a residual action of $G$. Reductions at general levels give Kostant varieties, which are closures
of complex coadjoint orbits.

This is universal in the following sense. Given a complex-symplectic space $M$ with a $G$ action, we may form its implosion 
\begin{equation}
M_{\rm impl} = (M \times \uri) /\!/ \Delta G
\end{equation}
which now has a $T$ action.  In terms of the Moore-Tachikawa category (enlarged to allow reductive groups), the universal 
implosion is an element of $\operatorname{Hom}(G,T)$ and a space with a $G$ action is viewed as living in $\operatorname{Hom}(1,G)$. The implosion operation
is now just composition with the universal implosion to obtain an element $M_{\rm impl}$ of $\operatorname{Hom}(1,T)$.

We may think of the universal right implosion $\uri$ as being obtained from $T^*G$, which has a $G \times G$ action, by breaking
the right action to a $T$ action.  Explicitly
\begin{equation}
\uri = \{ (g, v) \in G \times {\un}^{\circ}\} /\!/ U
\end{equation}
where $U$ denotes the maximal unipotent subgroup of $G$ acting by $(g,v) \mapsto (gu^{-1}, Ad(u)v)$, and
$\un = {\rm Lie}(U)$. Equivalently, the right implosion is the complex-symplectic quotient 
(in the Geometric Invariant Theory sense) of $T^*G$ by the right action of the maximal unipotent $U$. As a morphism in the enlarged Moore-Tachikawa category we have:
\begin{equation}\tag{\text{Universal Implosion}}
    \begin{tikzpicture}
        \node (1) at (0,0) {$G$};
        \node (2) at (2,0) {$T$.};
        \draw[thick,->] (1)--(2);
        \node at (1,0.3) {$\tensor{\operatorname{Q}}{_G}$};
    \end{tikzpicture}
\end{equation}
We also have an equivalent notion, denoted by \!\!$\uli$, of left implosion with a $T \times G$ action, so representing an element of $\operatorname{Hom}(T,G)$.
In the $G=\SL_{n}$ case we have an explicit quiver description of the implosion \cite{dancer2013implosion}.

A special feature of the implosion in complex symplectic geometry (not present for implosion in real symplectic geometry) is the presence of an extra symmetry by the Weyl group $W_G$ of $G$: there is an action on $Q_G$ of $G
\times (W_G\ltimes T)$.  A trace of this can be seen on the total space of the Grothendieck-Springer resolution of $G$, which is the quotient of an open subvariety of $Q_G$ by $T$.  The Weyl group only acts on the regular part of the Grothendieck-Springer resolution though, whereas it acts on all of $Q_G$.

In \cite{Dancer:2020wll} magnetic quivers for the implosion were introduced.
Let $G=\SL_{n+1}$ (so $T=\GL_1^n$), then, denoting as above the Coulomb branch by $\mathcal{C}$, we have:
\begin{equation}
    \uri=\mathcal{C}\left(\raisebox{-.5\height}{\begin{tikzpicture}
        \node[gauge,label=below:{$1$}] (1) at (1,0) {};
        \node[gauge,label=below:{$2$}] (2) at (2,0) {};
        \node (3) at (3,0) {$\cdots$};
        \node[gauge,label=below:{$n-1$}] (4) at (4,0) {};
        \node[gauge,label=below:{$n$}] (5) at (5,0) {};
        \draw (1)--(2)--(3)--(4)--(5);
        \node[gauge,label=right:{$1$}] (6u) at (6,1.5) {};
        \node at (6,0.5) {$\vdots$};
        \node[gauge,label=right:{$1$}] (6m) at (6,-0.5) {};
        \node[gauge,label=right:{$1$}] (6d) at (6,-1.5) {};
        \draw (5)--(6u) (5)--(6m) (5)--(6d);
        \draw [decorate,decoration={brace,amplitude=5pt},xshift=0,yshift=0]
    (6.5,1.8)--(6.5,-1.8) node [black,midway,xshift=0.7cm,yshift=0] {$n+1$};
    \end{tikzpicture}}\right)\;.
\end{equation}
The flavour symmetry of $X= \tensor{\operatorname{Q}}{_{\SL_{n+1}}}$ is $F_X=\SU(n+1)\times \Un(1)^n$. Gannon and Williams \cite{GW} have recently 
given an algebro-geometric proof that this is the correct magnetic quiver,
and observed that the Weyl group action on the implosion is manifest
as the symmetry permuting the Abelian bouquet nodes emanating from the
$n$-dimensional node. It is interesting to note that the Weyl action is much less obvious in the quiver Higgs branch description of the implosion given in 
\cite{dancer2013implosion} (for an explicit description of the action in this picture see Wang \cite{wang:2021}).

\paragraph{Partial Implosion} We also have a notion of partial implosions, where we take the complex-symplectic quotient in the GIT sense of $T^*G$ by the maximal
unipotent $U_P$ of a parabolic subgroup $P$ of $G$. The universal contraction is
\begin{equation}
\urip=(G \times {\un}_{P}^\circ)/\!/U_P.
\end{equation}
This has an action of $G \times L_P$ where $L_P$ is the Levi subgroup of $P$, as $P$ is the semidirect product
of $U_P$ and $L_P$, so $L_P$ normalises $U_P$.
The case when $P$ is the Borel recovers the ordinary implosion. In general
we have:

\begin{equation}\tag{Partial Implosion}
    \begin{tikzpicture}
        \node (1) at (0,0) {$G$};
        \node (2) at (2,0) {$L_P$.};
        \draw[thick,->] (1)--(2);
        \node at (1,0.3) {$\urip$};
    \end{tikzpicture}\;
\end{equation}
where $L_P$ is a Levi of $G$. 

When $G=\SL_{n+1}$ and $L_P=S(\prod_{i=1}^l\GL_{k_i})$ (where $\sum_{i=1}^lk_i=n+1$ and $k_1\geq\dots\geq k_l>0$),
\begin{equation}
    \urip=\mathcal{C}\left(\raisebox{-.5\height}{\begin{tikzpicture}
        \node[gauge,label=below:{$1$}] (1) at (1,0) {};
        \node[gauge,label=below:{$2$}] (2) at (2,0) {};
        \node (3) at (3,0) {$\cdots$};
        \node[gauge,label=below:{$n-1$}] (4) at (4,0) {};
        \node[gauge,label=below:{$n$}] (5) at (5,0) {};
        \draw (1)--(2)--(3)--(4)--(5);
        \node[gauge,label=above:{$k_1$}] (6u) at (6,1.5) {};
        \node (7u) at (7,1.5) {$\cdots$};
        \node[gauge,label=above:{$1$}] (8u) at (8,1.5) {};
        \node at (6,0.5) {$\vdots$};
        \node[gauge,label=above:{$k_{l-1}$}] (6m) at (6,-0.5) {};
        \node (7m) at (7,-0.5) {$\cdots$};
        \node[gauge,label=above:{$1$}] (8m) at (8,-0.5) {};
        \node[gauge,label=below:{$k_l$}] (6d) at (6,-1.5) {};
        \node (7d) at (7,-1.5) {$\cdots$};
        \node[gauge,label=below:{$1$}] (8d) at (8,-1.5) {};
        \draw (5)--(6u)--(7u)--(8u) (5)--(6m)--(7m)--(8m) (5)--(6d)--(7d)--(8d);
    \end{tikzpicture}}\right)\;.
\end{equation}
$\urip$ is singular and the flavour symmetry of $\urip$ is $F_{\urip}=\SU(n+1)\times S(\prod_{i=1}^l \Un(k_i))$, unless $k_1=n$ when $\urip=\mathbb{H}^{n(n+1)}$ and $F_{\urip}=Sp(n(n+1))$, or $k_1=n+1$ when $\urip=T^*\SL_{n+1}$ and $F_{\urip}=\SU(n+1)^2$.

\paragraph{Universal Contraction}
The universal contraction space corresponds to composing the left and right implosion morphisms in the enlarged
Moore-Tachikawa category
\begin{equation}
    \begin{tikzpicture}
        \node (1) at (0,0) {$G$};
        \node (2) at (2,0) {$T$};
        \node (3) at (4,0) {$G$,};
        \draw[thick,->] (1)--(2);
        \draw[thick,->] (2)--(3);
        \node at (1,0.3) {${\tensor{\operatorname{Q}}{_G}}$};
        \node at (3,0.3) {${\uli}$};
        \draw[thick,->] (1) .. controls (2,-1) .. (3);
        \node at (2,-1.1) {$\ucon
        =\uli\circ \tensor{\operatorname{Q}}{_G}$};
    \end{tikzpicture}\;
\end{equation}
so that we pick up an extra $T$ action as the middle group is Abelian (in fact, we also keep the action of one copy of the Weyl group $W_G$).

Equivalently we form the complex-symplectic reduction
\begin{equation}
\ucon=(\uli \times \tensor{\operatorname{Q}}{_G})/\!/_0 T,
\end{equation}
which now has a $G \times \left(W_G\ltimes T\right) \times G$ action. The dimension is the same as $T^*G$ but we have enhanced the 
symmetry group with an extra $T$ factor. (In \cite{2024arXiv240107920D} we referred to $\ucon$
as $(T^*G)_{\rm csc}$.)

Note that the complex-symplectic reduction at level zero of the universal contraction $\ucon$ by $T$ is the product
of two copies of the reduction of the implosion by $T$, i.e.\ the product of two copies of the nilpotent cone. We shall use this to check our conjectures for the Coulomb branches.

Let $G=SL_{n+1}$, then
\begin{equation}
    \ucon=\uli\circ \uri=\mathcal{C}\left(\raisebox{-.5\height}{\begin{tikzpicture}
        \node[gauge,label=below:{$1$}] (1) at (1,0) {};
        \node[gauge,label=below:{$2$}] (2) at (2,0) {};
        \node (3) at (3,0) {$\cdots$};
        \node[gauge,label=below:{$n-1$}] (4) at (4,0) {};
        \node[gauge,label=below:{$n$}] (5) at (5,0) {};
        \draw (1)--(2)--(3)--(4)--(5);
        \node[gauge,label=below:{$1$}] (11) at (11,0) {};
        \node[gauge,label=below:{$2$}] (10) at (10,0) {};
        \node (9) at (9,0) {$\cdots$};
        \node[gauge,label=below:{$n-1$}] (8) at (8,0) {};
        \node[gauge,label=below:{$n$}] (7) at (7,0) {};
        \draw (7)--(8)--(9)--(10)--(11);
        \node[gauge,label=above:{$1$}] (6u) at (6,1.5) {};
        \node at (6,0.5) {$\vdots$};
        \node[gauge,label=below:{$1$}] (6m) at (6,-0.5) {};
        \node[gauge,label=below:{$1$}] (6d) at (6,-1.5) {};
        \draw (5)--(6u)--(7) (5)--(6m)--(7) (5)--(6d)--(7);
    \end{tikzpicture}}\right)\;.
    \label{2.11}
\end{equation}
The flavour symmetry of $X=\ucon$ is $F_X=\SU(n+1)\times \Un(1)^n\times \SU(n+1)$. This is a special case of the partial contraction.

The reduction of $\ucon$ by $T$ can be realised on the magnetic quiver by turning the U$(1)$ gauge nodes in the middle into flavour nodes, giving
\begin{equation}
    \begin{split}
        \ucon/\!/\!/T=&\mathcal{C}\left(\raisebox{-.5\height}{\begin{tikzpicture}
        \node[gauge,label=below:{$1$}] (1) at (1,0) {};
        \node[gauge,label=below:{$2$}] (2) at (2,0) {};
        \node (3) at (3,0) {$\cdots$};
        \node[gauge,label=below:{$n-1$}] (4) at (4,0) {};
        \node[gauge,label=below:{$n$}] (5) at (5,0) {};
        \draw (1)--(2)--(3)--(4)--(5);
        \node[gauge,label=below:{$1$}] (11) at (11,0) {};
        \node[gauge,label=below:{$2$}] (10) at (10,0) {};
        \node (9) at (9,0) {$\cdots$};
        \node[gauge,label=below:{$n-1$}] (8) at (8,0) {};
        \node[gauge,label=below:{$n$}] (7) at (7,0) {};
        \draw (7)--(8)--(9)--(10)--(11);
        \node[flavour,label=above:{$1$}] (6u) at (6,1.5) {};
        \node at (6,0.5) {$\vdots$};
        \node[flavour,label=below:{$1$}] (6m) at (6,-0.5) {};
        \node[flavour,label=below:{$1$}] (6d) at (6,-1.5) {};
        \draw (5)--(6u)--(7) (5)--(6m)--(7) (5)--(6d)--(7);
    \end{tikzpicture}}\right)\\
    =&\mathcal{C}\left(\raisebox{-.5\height}{\begin{tikzpicture}
        \node[gauge,label=below:{$1$}] (1) at (1,0) {};
        \node[gauge,label=below:{$2$}] (2) at (2,0) {};
        \node (3) at (3,0) {$\cdots$};
        \node[gauge,label=below:{$n-1$}] (4) at (4,0) {};
        \node[gauge,label=below:{$n$}] (5) at (5,0) {};
        \draw (1)--(2)--(3)--(4)--(5);
        \node[gauge,label=below:{$1$}] (11) at (11,0) {};
        \node[gauge,label=below:{$2$}] (10) at (10,0) {};
        \node (9) at (9,0) {$\cdots$};
        \node[gauge,label=below:{$n-1$}] (8) at (8,0) {};
        \node[gauge,label=below:{$n$}] (7) at (7,0) {};
        \draw (7)--(8)--(9)--(10)--(11);
        \node[flavour,label=above:{$n+1$}] (6) at (6,0) {};
        \draw (5)--(6)--(7);
    \end{tikzpicture}}\right)=\mathcal{N}^2\;,
    \end{split}
    \label{2.12}
\end{equation}
where $\mathcal{N}$ is the nilpotent cone of $\SL_{n+1}$, as expected.

\paragraph{Partial Contraction}
We now consider the composition
\begin{equation}
    \begin{tikzpicture}
        \node (1) at (0,0) {$G$};
        \node (2) at (2,0) {$L$};
        \node (3) at (4,0) {$G$.};
        \draw[thick,->] (1)--(2);
        \draw[thick,->] (2)--(3);
        \node at (1,0.3) {$\ulip$};
        \node at (3,0.3) {$\urip$};
        \draw[thick,->] (1) .. controls (2,-1) .. (3);
        \node at (2,-1.1) {$\uconp=\urip\circ \ulip$};
    \end{tikzpicture}\;
\end{equation}
That is, we take the product of the partial implosions and symplectically reduce by the Levi.
We now obtain only an action of $G \times Z(L) \times G$ where $Z(L)$ is the centre of the Levi.
Let $G=SL_{n+1}$ and $L=S(\prod_{i=1}^{l}\GL_{k_i})$, where $\sum_{i=1}^{l}k_i=n+1$, so $Z(L)$ is the $(l-1)$-dimensional torus. We propose
\begin{equation}
    \uconp=\ulip\circ \urip=\mathcal{C}\left(\raisebox{-.5\height}{\begin{tikzpicture}
        \node[gauge,label=below:{$1$}] (1) at (1,0) {};
        \node[gauge,label=below:{$2$}] (2) at (2,0) {};
        \node (3) at (3,0) {$\cdots$};
        \node[gauge,label=below:{$n-1$}] (4) at (4,0) {};
        \node[gauge,label=below:{$n$}] (5) at (5,0) {};
        \draw (1)--(2)--(3)--(4)--(5);
        \node[gauge,label=below:{$1$}] (11) at (11,0) {};
        \node[gauge,label=below:{$2$}] (10) at (10,0) {};
        \node (9) at (9,0) {$\cdots$};
        \node[gauge,label=below:{$n-1$}] (8) at (8,0) {};
        \node[gauge,label=below:{$n$}] (7) at (7,0) {};
        \draw (7)--(8)--(9)--(10)--(11);
        \node[gauge,label=above:{$k_1$}] (6u) at (6,1.5) {};
        \node at (6,0.5) {$\vdots$};
        \node[gauge,label=above:{$k_{l-1}$}] (6m) at (6,-0.5) {};
        \node[gauge,label=below:{$k_l$}] (6d) at (6,-1.5) {};
        \draw (5)--(6u)--(7) (5)--(6m)--(7) (5)--(6d)--(7);
    \end{tikzpicture}}\right)\;.
\end{equation}
The flavour symmetry of $\uconp$ is $F_{\uconp}=\SU(n+1)\times \Un(1)^{(l-1)}\times \SU(n+1)$.

The reduction by $\Un(1)^{(l-1)}$ is
\begin{equation}
    \uconp/\!/\!/\Un(1)^l=\mathcal{C}\left(\raisebox{-.5\height}{\begin{tikzpicture}
        \node[gauge,label=below:{$1$}] (1) at (1,0) {};
        \node[gauge,label=below:{$2$}] (2) at (2,0) {};
        \node (3) at (3,0) {$\cdots$};
        \node[gauge,label=below:{$n-1$}] (4) at (4,0) {};
        \node[gauge,label=below:{$n$}] (5) at (5,0) {};
        \draw (1)--(2)--(3)--(4)--(5);
        \node[gauge,label=below:{$1$}] (11) at (11,0) {};
        \node[gauge,label=below:{$2$}] (10) at (10,0) {};
        \node (9) at (9,0) {$\cdots$};
        \node[gauge,label=below:{$n-1$}] (8) at (8,0) {};
        \node[gauge,label=below:{$n$}] (7) at (7,0) {};
        \draw (7)--(8)--(9)--(10)--(11);
        \node[gauge,label=above:{\scriptsize$\SU(k_1)$}] (6u) at (6,1.5) {};
        \node at (6,0.5) {$\vdots$};
        \node[gauge,label=above:{\scriptsize$\SU(k_{l-1})$}] (6m) at (6,-0.5) {};
        \node[gauge,label=below:{\scriptsize$\SU(k_l)$}] (6d) at (6,-1.5) {};
        \draw (5)--(6u)--(7) (5)--(6m)--(7) (5)--(6d)--(7);
        \draw (11,-1)--(12,0);
        \node at (12,-1) {$\mathbb{Z}_{\mathrm{gcd(k_i)}}$};
    \end{tikzpicture}}\right)\;,
    \label{mirrorpair}
\end{equation}
where the notation $/\mathbb{Z}_{\mathrm{gcd(k_i)}}$ denotes the gauging of 1-form symmetry.

It was shown in \cite{jia:2021,gannon2024proof} that the universal implosions have symplectic singularities.  We expect the same to hold for the universal contractions.

\section{3d Mirrors}
\label{sec:2}

In \cite{Bourget:2021qpx}, the 3d mirrors of implosions of quivers were discussed.

A standard example of a 3d mirror pair (in fact an example of a self-mirror) is the following
\begin{equation}
    \begin{tikzpicture}
        \node (a) at (0,0) {$T[\SU(n)]=\raisebox{-.5\height}{\begin{tikzpicture}
        \node[gauge,label=below:{$1$}] (1) at (1,0) {};
        \node[gauge,label=below:{$2$}] (2) at (2,0) {};
        \node (3) at (3,0) {$\cdots$};
        \node[gauge,label=below:{$n-1$}] (4) at (4,0) {};
        \node[gauge,label=below:{$n$}] (5) at (5,0) {};
        \draw (1)--(2)--(3)--(4)--(5);
        \node[flavour,label=below:{$n+1$}] (6) at (6,0) {};
        \draw (5)--(6);
    \end{tikzpicture}}$};
        \node (b) at (9,0) {$\raisebox{-.5\height}{\begin{tikzpicture}
        \node[gauge,label=below:{$1$.}] (1) at (6,0) {};
        \node[gauge,label=below:{$2$}] (2) at (5,0) {};
        \node (3) at (4,0) {$\cdots$};
        \node[gauge,label=below:{$n-1$}] (4) at (3,0) {};
        \node[gauge,label=below:{$n$}] (5) at (2,0) {};
        \draw (1)--(2)--(3)--(4)--(5);
        \node[flavour,label=below:{$n+1$}] (6) at (1,0) {};
        \draw (5)--(6);
    \end{tikzpicture}}$};
    \draw [<->] (a) -- node [midway,above,align=center] {\scriptsize 3d MS} (b);
    \end{tikzpicture}\;
\end{equation}
We have the following 3d mirror pair involving the universal implosion quiver:
\begin{equation}
    \begin{tikzpicture}
        \node (a) at (0,0) {$\raisebox{-.5\height}{\begin{tikzpicture}
        \node[gauge,label=below:{$1$}] (1) at (1,0) {};
        \node[gauge,label=below:{$2$}] (2) at (2,0) {};
        \node (3) at (3,0) {$\cdots$};
        \node[gauge,label=below:{$n-1$}] (4) at (4,0) {};
        \node[gauge,label=below:{$n$}] (5) at (5,0) {};
        \draw (1)--(2)--(3)--(4)--(5);
        \node[gauge,label=right:{$1$}] (6u) at (6,1.5) {};
        \node at (6,0.5) {$\vdots$};
        \node[gauge,label=right:{$1$}] (6m) at (6,-0.5) {};
        \node[gauge,label=right:{$1$}] (6d) at (6,-1.5) {};
        \draw (5)--(6u) (5)--(6m) (5)--(6d);
    \end{tikzpicture}}$};
        \node (b) at (8,0) {$\raisebox{-.5\height}{\begin{tikzpicture}
        \node[gauge,label=below:{{\tiny$SU(1)$},}] (1) at (6,0) {};
        \node[gauge,label=below:{\tiny$SU(2)$}] (2) at (5,0) {};
        \node (3) at (4,0) {$\cdots$};
        \node[gauge,label=below:{\tiny$SU(n-1)$}] (4) at (3,0) {};
        \node[gauge,label=below:{\tiny$SU(n)$}] (5) at (2,0) {};
        \draw (1)--(2)--(3)--(4)--(5);
        \node[flavour,label=below:{\tiny$n+1$}] (6) at (1,0) {};
        \draw (5)--(6);
    \end{tikzpicture}}$};
    \draw [<->] (a) -- node [midway,above,align=center] {\scriptsize 3d MS} (b);
    \end{tikzpicture}\;
\end{equation}
where the theory on the left is $T[\SU(n)]$, but with the $n+1$ flavour node `exploded' into $n+1$ U$(1)$ gauge nodes, i.e.\ we gauge the U$(1)^n$ torus inside the flavour symmetry of the Higgs branch; and the theory on the right is $T[\SU(n)]$, but with all the unitary gauge groups replaced with special unitary gauge groups, i.e.\ we gauge the U$(1)^n$ torus inside the flavour symmetry of the Coulomb branch. 

The universal contraction quiver can be obtained by taking two copies of $T[\SU(n)]$ and gauging the diagonal $\Un(1)^n$ of the tori of their Higgs branches. 
Therefore we expect that the 3d mirror is obtained by taking two copies of $T[\SU(n)]$ and gauging the diagonal $\Un(1)^n$ of the tori of their Coulomb branches.
\begin{equation}
\label{eq:Contraction3dMirror}
    \begin{tikzpicture}
        \node (a) at (0,0) {$\raisebox{-.5\height}{\begin{tikzpicture}
        \node[gauge,label=below:{$1$}] (1) at (1,0) {};
        \node[gauge,label=below:{$2$}] (2) at (2,0) {};
        \node (3) at (3,0) {$\cdots$};
        \node[gauge,label=below:{$n-1$}] (4) at (4,0) {};
        \node[gauge,label=below:{$n$}] (5) at (5,0) {};
        \draw (1)--(2)--(3)--(4)--(5);
        \node[gauge,label=below:{$1$}] (11) at (11,0) {};
        \node[gauge,label=below:{$2$}] (10) at (10,0) {};
        \node (9) at (9,0) {$\cdots$};
        \node[gauge,label=below:{$n-1$}] (8) at (8,0) {};
        \node[gauge,label=below:{$n$}] (7) at (7,0) {};
        \draw (7)--(8)--(9)--(10)--(11);
        \node[gauge,label=above:{$1$}] (6u) at (6,1.5) {};
        \node at (6,0.5) {$\vdots$};
        \node[gauge,label=below:{$1$}] (6m) at (6,-0.5) {};
        \node[gauge,label=below:{$1$}] (6d) at (6,-1.5) {};
        \draw (5)--(6u)--(7) (5)--(6m)--(7) (5)--(6d)--(7);
    \end{tikzpicture}}$};
        \node (b) at (0,-5) {$\raisebox{-.5\height}{\begin{tikzpicture}
        \node[gauge,label=below:{\tiny${\color{red}S[}U(1)$}] (1) at (6,0) {};
        \node[gauge,label=below:{\tiny${\color{blue}S[}U(2)$}] (2) at (5,0) {};
        \node (3) at (4,0) {$\cdots$};
        \node[gauge,label=below:{\tiny${\color{green}S[}U(n-1)$}] (4) at (3,0) {};
        \node[gauge,label=below:{\tiny${\color{orange}S[}U(n)$}] (5) at (2,0) {};
        \draw (1)--(2)--(3)--(4)--(5);
        \node[flavour,label=below:{\tiny$n+1$}] (6) at (1,0) {};
        \draw (5)--(6);
        \node[gauge,label=above:{\tiny$U(1){\color{red}]}$}] (1d) at (6,-1.5) {};
        \node[gauge,label=above:{\tiny$U(2){\color{blue}]}$}] (2d) at (5,-1.5) {};
        \node (3d) at (4,-1.5) {$\cdots$};
        \node[gauge,label=above:{\tiny$U(n-1){\color{green}]}$}] (4d) at (3,-1.5) {};
        \node[gauge,label=above:{\tiny$U(n){\color{orange}]}$}] (5d) at (2,-1.5) {};
        \draw (1d)--(2d)--(3d)--(4d)--(5d);
        \node[flavour,label=above:{\tiny$n+1$}] (6d) at (1,-1.5) {};
        \draw (5d)--(6d);
    \end{tikzpicture}}$};
    \draw [<->] (a) -- node [midway,right,align=center] {\scriptsize 3d MS} (b);
    \end{tikzpicture}
\end{equation}

The colours represent gauge groups where a diagonal $\Un(1)$ is decoupled from them. This result has been verified through explicit computation of Hilbert series of both quivers. 

The more non-trivial side of the 3d mirror is to check the Higgs branch of the contracted quiver with the Coulomb branch of the product of two linear quivers. To do this, we compute the Hilbert series for the two coordinate rings (equivalently, the Higgs and Coulomb branch chiral ring from a physics point of view). 

Note that in the contracted quiver, although there are no flavour nodes, we expect a $U(1)^{n-1}$ symmetry from the loops in the graph. We can compare this with the
case of Asymptotically Locally Euclidean hyper-K\"ahler manifolds of quaternionic dimension one, which are known to arise from the extended Dynkin diagrams of $ADE$ type. In the $A_n$ case,
uniquely, the extended diagram is a loop and this gives rise to the $U(1)$-triholomorphic symmetry
that occurs in this case but not in the $D_n$ or $E_n$ cases where the extended diagrams are trees.

The corresponding deformation parameters in the Coulomb branch are the choice of level set in the description of the contraction as the hyper-K\"ahler quotient
by $T \cong \Un(1)^{n-1}$ of the product of two implosions.

For $n=3$, the resulting Hilbert series is:
\begin{equation}
\begin{split}
   HS^{\text{Contraction}}_{\text{Higgs}}(t)&=\frac{1+6 t^4+12 t^6+18 t^8+24 t^{10}+34 t^{12}+24 t^{14}+18 t^{16}+12 t^{18}+6
   t^{20}+t^{24}}{(-1+t)^8 (1+t)^8 \left(1+t^2\right)^4 \left(1-t+t^2\right)^2
   \left(1+t+t^2\right)^2} \\
   &=1+2 t^2+13 t^4+38 t^6+115 t^8+284 t^{10}+666 t^{12}+1392 t^{14}+O\left(t^{15}\right)
   \end{split}
     \label{HiggsHS3}
\end{equation}
\\
which is the same as the Coulomb branch Hilbert series of the product quivers at the bottom of \eqref{eq:Contraction3dMirror}.
In the second line, we expressed the result as a Taylor expansion in $t$. The coefficient of the leading term $2t^2$ is in accordance with our observation above of the $\Un(1)^2$ Higgs branch global
symmetry group of the contracted quiver. The exact equivalence of the two Hilbert series gives very strong indication that the two moduli spaces are the same. 

The other side of the duality can also be checked, but due to the complication of the computation, we only provide the first few terms:
\begin{equation}
    HS^{\text{Contraction}}_{\text{Coulomb}}(t)=1+18 t^2+223 t^4+1966 t^6+13444 t^8+74590 t^{10}+O\left(t^{12}\right).
    \label{CoulombHS3}
\end{equation}
This is the same with the Higgs branch Hilbert series of the product quivers at the bottom of \eqref{eq:Contraction3dMirror}. The coefficient of the leading term $18t^2$ is consistent with the expectation that Coulomb branch global symmetry group of the contracted quivers is $\SU(3)^2\times \Un(1)^2$. The computations \eqref{HiggsHS3} and \eqref{CoulombHS3}  give strong indications that the mirror pair in \eqref{mirrorpair} is correct for $n=3$. Recall \cite{dancer2013implosion},\cite{Dancer:2020wll} that uniquely in the $n=3$ case the symmetry
group of the implosion is enhanced from $SU(3) \times \Un(1)^2$
to $SO(8)$, reflecting the fact that the Abelian bouquet nodes are balanced in this case. However when we form the contraction only the
$SU(3)^2 \times \Un(1)^2$ survives. In terms of the quiver diagram, the bouquet nodes become unbalanced again when we fuse the implosion
quivers together.

For $n=4$, the resulting Hilbert series is:
\begin{equation}
\begin{split}
   HS^{\text{Contraction}}_{\text{Higgs}}(t)&=\frac{\parbox{11cm}{$1-t^2+16 t^4+41 t^6+164 t^8+499 t^{10}+1437 t^{12}+3336 t^{14}+7492 t^{16}+14524
   t^{18}+26484 t^{20}+43272 t^{22}+66107 t^{24}+91941 t^{26}+119980 t^{28}+143407
   t^{30}+160831 t^{32}+166042 t^{34}+160831 t^{36}+143407 t^{38}+119980 t^{40}+91941
   t^{42}+66107 t^{44}+43272 t^{46}+26484 t^{48}+14524 t^{50}+7492 t^{52}+3336 t^{54}+1437
   t^{56}+499 t^{58}+164 t^{60}+41 t^{62}+16 t^{64}-t^{66}+t^{68}$}}{(-1+t)^{18} (1+t)^{18}
   \left(1+t^2\right)^9 \left(1-t+t^2\right)^5 \left(1+t+t^2\right)^5
   \left(1+t^4\right)^3} \\
   &=1+3 t^2+28 t^4+138 t^6+674 t^8+2808 t^{10}+10781 t^{12}+37479 t^{14}+O\left(t^{15}\right)
   \end{split}
     \label{HiggsHS4}
\end{equation}
which is the same as the Coulomb branch Hilbert series of the product quivers.

The other side of the duality can also be checked with the first few terms being:
\begin{equation}
    HS^{\text{Contraction}}_{\text{Coulomb}}(t)=1+33t^2+559t^4+6685t^6+61900t^8+491584t^{10}+O\left(t^{12}\right).
    \label{CoulombHS4}
\end{equation}
Again, we see the $t^2$ coefficient gives the dimension of the global symmetry group
$SU(4)^2 \times \Un(1)^3$ of the contraction.

Note that, like in the case of implosion, the action of the Weyl group is not reflected in the Hilbert series, related to the fact that it does not respect the $\mathbb{C}^*$-action on the universal implosion or contraction spaces.

\paragraph{Implications for 3d mirror pairs.}
The results of this section have interesting implications. The study of linear quivers\footnote{Here, linear quivers are framed quivers whose gauge groups link together to form a line. Otherwise, they are non-linear (e.g. quivers with the shape of D-type Dynkin diagrams.) } with mixed unitary and special unitary gauge groups -- and in particular the rules to derive their mirrors -- is explored in great detail in \cite{Bourget:2021jwo}. These 3d mirrors cover a large family of quivers with a bouquet of $U(1)$ gauge nodes, however quivers like \eqref{2.11} do not appear.  Whilst \cite{Bourget:2021jwo} studies changing single unitary gauge nodes to special unitary gauge nodes (via gauging the topological $\Un(1)$ symmetry of the unitary node), it never investigates the effect of gauging the diagonal $\Un(1)$ inside the topological $\Un(1)^k$ symmetry associated to $k$ unitary nodes, e.g.\ $S[\prod_{i=1}^k \Un(n_i)]$. Here we showed that it is precisely this diagonal gauging of a linear unitary quiver, which leads to the non-linear 3d mirror quiver \eqref{2.11}. 

The process of gauging diagonal $\Un(1)$s sheds light on finding mirrors of non-linear quivers and expanding the landscape of 3d mirror pairs. For example, rather than taking two copies of $T[SU(n)]$ theories and gauging the $U(1)^{n-1}$ subgroup of the $SU(n)$ flavor symmetry as we've done in \eqref{eq:Contraction3dMirror}, consider three pairs instead: 
\begin{equation}
\scalebox{0.9}{
\begin{tikzpicture}
	\begin{pgfonlayer}{nodelayer}
		\node [style=gauge3] (0) at (-1.75, -2.25) {};
		\node [style=gauge3] (1) at (-0.75, -2.25) {};
		\node [style=gauge3] (2) at (1.25, -2.25) {};
		\node [style=flavour2] (3) at (2.25, -2.25) {};
		\node [style=none] (4) at (-0.25, -2.25) {};
		\node [style=none] (5) at (0.75, -2.25) {};
		\node [style=none] (6) at (0.25, -2.25) {$\dots$};
		\node [style=none] (7) at (-1.75, -2.75) {1};
		\node [style=none] (8) at (-0.75, -2.75) {2};
		\node [style=none] (9) at (1.25, -2.75) {$n-1$};
		\node [style=none] (10) at (2.25, -2.75) {$n$};
		\node [style=gauge3] (11) at (7.25, -2.25) {};
		\node [style=gauge3] (12) at (6.25, -2.25) {};
		\node [style=gauge3] (13) at (4.25, -2.25) {};
		\node [style=flavour2] (14) at (3.25, -2.25) {};
		\node [style=none] (15) at (5.75, -2.25) {};
		\node [style=none] (16) at (4.75, -2.25) {};
		\node [style=none] (17) at (5.25, -2.25) {$\dots$};
		\node [style=none] (18) at (7.25, -2.75) {1};
		\node [style=none] (19) at (6.25, -2.75) {2};
		\node [style=none] (20) at (4.25, -2.75) {$n-1$};
		\node [style=none] (21) at (3.25, -2.75) {$n$};
		\node [style=gauge3] (22) at (2.75, 2.5) {};
		\node [style=gauge3] (23) at (2.75, 1.5) {};
		\node [style=gauge3] (24) at (2.75, -0.5) {};
		\node [style=flavour2] (25) at (2.75, -1.5) {};
		\node [style=none] (26) at (2.75, 1) {};
		\node [style=none] (27) at (2.75, 0) {};
		\node [style=none] (28) at (2.75, 0.5) {$\vdots$};
		\node [style=none] (29) at (3.25, 2.5) {1};
		\node [style=none] (30) at (3.25, 1.5) {2};
		\node [style=none] (31) at (3.5, -0.5) {$n-1$};
		\node [style=none] (32) at (3.25, -1.5) {$n$};
		\node [style=none] (33) at (2.75, -1) {};
		\node [style=none] (34) at (1.75, -2) {};
		\node [style=none] (35) at (3.75, -2) {};
		\node [style=none] (36) at (2.75, -3) {};
		\node [style=none] (37) at (2.75, -3.5) {\textcolor{brown}{Gauge $U(1)^{n-1}\subset SU(n)$}};
		\node [style=none] (38) at (8, -11.5) {};
		\node [style=none] (39) at (9.5, -11.5) {};
		\node [style=none] (41) at (8.75, -11) {3d mirror};
		\node [style=gauge3] (42) at (10, -9.25) {};
		\node [style=gauge3] (43) at (11.25, -9.25) {};
		\node [style=gauge3] (44) at (13.25, -9.25) {};
		\node [style=flavour2] (45) at (14.25, -9.25) {};
		\node [style=none] (46) at (11.75, -9.25) {};
		\node [style=none] (47) at (12.75, -9.25) {};
		\node [style=none] (48) at (12.25, -9.25) {$\dots$};
		\node [style=none] (49) at (10, -9.75) {\textcolor{cyan}{S[U(1)}};
		\node [style=none] (52) at (14.25, -9.75) {$n$};
		\node [style=gauge3] (53) at (10, -10.75) {};
		\node [style=gauge3] (54) at (11.25, -10.75) {};
		\node [style=gauge3] (55) at (13.25, -10.75) {};
		\node [style=flavour2] (56) at (14.25, -10.75) {};
		\node [style=none] (57) at (11.75, -10.75) {};
		\node [style=none] (58) at (12.75, -10.75) {};
		\node [style=none] (59) at (12.25, -10.75) {$\dots$};
		\node [style=none] (60) at (10, -11.25) {\textcolor{cyan}{U(1)}};
		\node [style=none] (63) at (14.25, -11.25) {$n$};
		\node [style=gauge3] (64) at (10, -12.5) {};
		\node [style=gauge3] (65) at (11.25, -12.5) {};
		\node [style=gauge3] (66) at (13.25, -12.5) {};
		\node [style=flavour2] (67) at (14.25, -12.5) {};
		\node [style=none] (68) at (11.75, -12.5) {};
		\node [style=none] (69) at (12.75, -12.5) {};
		\node [style=none] (70) at (12.25, -12.5) {$\dots$};
		\node [style=none] (71) at (10, -13) {\textcolor{cyan}{U(1)]}};
		\node [style=none] (74) at (14.25, -13) {$n$};
		\node [style=none] (75) at (11.25, -9.75) {\textcolor{red}{S[U(2)}};
		\node [style=none] (76) at (11.25, -11.25) {\textcolor{red}{U(2)}};
		\node [style=none] (77) at (11.25, -13) {\textcolor{red}{U(2)]}};
		\node [style=none] (78) at (13.25, -9.75) {\textcolor{blue}{S[U(n-1)}};
		\node [style=none] (79) at (13.25, -11.25) {\textcolor{blue}{U(n-1)}};
		\node [style=none] (80) at (13.25, -13) {\textcolor{blue}{U(n-1)]}};
		\node [style=gauge3] (81) at (-1.75, -11) {};
		\node [style=gauge3] (82) at (-0.75, -11) {};
		\node [style=gauge3] (83) at (1.25, -11) {};
		\node [style=none] (85) at (-0.25, -11) {};
		\node [style=none] (86) at (0.75, -11) {};
		\node [style=none] (87) at (0.25, -11) {$\dots$};
		\node [style=none] (88) at (-1.75, -11.5) {1};
		\node [style=none] (89) at (-0.75, -11.5) {2};
		\node [style=none] (90) at (1.25, -11.5) {$n-1$};
		\node [style=gauge3] (92) at (7.25, -11) {};
		\node [style=gauge3] (93) at (6.25, -11) {};
		\node [style=gauge3] (94) at (4.25, -11) {};
		\node [style=none] (96) at (5.75, -11) {};
		\node [style=none] (97) at (4.75, -11) {};
		\node [style=none] (98) at (5.25, -11) {$\dots$};
		\node [style=none] (99) at (7.25, -11.5) {1};
		\node [style=none] (100) at (6.25, -11.5) {2};
		\node [style=none] (101) at (4.25, -11.5) {$n-1$};
		\node [style=gauge3] (103) at (2.75, -5.5) {};
		\node [style=gauge3] (104) at (2.75, -6.5) {};
		\node [style=gauge3] (105) at (2.75, -8.5) {};
		\node [style=none] (107) at (2.75, -7) {};
		\node [style=none] (108) at (2.75, -8) {};
		\node [style=none] (109) at (2.75, -7.5) {$\vdots$};
		\node [style=none] (110) at (3.25, -5.5) {1};
		\node [style=none] (111) at (3.25, -6.5) {2};
		\node [style=none] (112) at (3.5, -8.5) {$n-1$};
		\node [style=none] (120) at (0.5, -1.5) {};
		\node [style=gauge3] (121) at (2.75, -9.5) {};
		\node [style=gauge3] (122) at (2.75, -10.25) {};
		\node [style=gauge3] (123) at (2.75, -11) {};
		\node [style=gauge3] (124) at (2.75, -12.5) {};
		\node [style=none] (127) at (2.75, -11.75) {$\vdots$};
		\node [style=none] (128) at (2.25, -9.5) {1};
		\node [style=none] (129) at (2.25, -10.25) {1};
		\node [style=none] (130) at (2.25, -11.25) {1};
		\node [style=none] (131) at (2.75, -13) {1};
		\node [style=none] (132) at (2.625, -4.25) {};
		\node [style=none] (133) at (2.625, -4.75) {};
		\node [style=none] (134) at (2.875, -4.75) {};
		\node [style=none] (135) at (2.875, -4.25) {};
	\end{pgfonlayer}
	\begin{pgfonlayer}{edgelayer}
		\draw (0) to (1);
		\draw (1) to (4.center);
		\draw (5.center) to (3);
		\draw (11) to (12);
		\draw (12) to (15.center);
		\draw (16.center) to (14);
		\draw (22) to (23);
		\draw (23) to (26.center);
		\draw (27.center) to (25);
		\draw [style=browndotted, bend right=45] (34.center) to (36.center);
		\draw [style=browndotted, bend right=45] (36.center) to (35.center);
		\draw [style=browndotted, bend right=45] (35.center) to (33.center);
		\draw [style=browndotted, bend left=315] (33.center) to (34.center);
		\draw [style=arrowed] (38.center) to (39.center);
		\draw [style=arrowed] (39.center) to (38.center);
		\draw (42) to (43);
		\draw (43) to (46.center);
		\draw (47.center) to (45);
		\draw (53) to (54);
		\draw (54) to (57.center);
		\draw (58.center) to (56);
		\draw (64) to (65);
		\draw (65) to (68.center);
		\draw (69.center) to (67);
		\draw (81) to (82);
		\draw (82) to (85.center);
		\draw (92) to (93);
		\draw (93) to (96.center);
		\draw (103) to (104);
		\draw (104) to (107.center);
		\draw (108.center) to (105);
		\draw (83) to (121);
		\draw (83) to (122);
		\draw (83) to (123);
		\draw (83) to (124);
		\draw (94) to (124);
		\draw (123) to (94);
		\draw (94) to (122);
		\draw (94) to (121);
		\draw (105) to (121);
		\draw [bend left] (105) to (122);
		\draw [bend right=45] (123) to (105);
		\draw [bend right=45] (124) to (105);
		\draw (135.center) to (134.center);
		\draw (133.center) to (132.center);
	\end{pgfonlayer}
\end{tikzpicture}
}
\end{equation}
This relation can be checked in the same way as we checked \eqref{eq:Contraction3dMirror} by computing the refined Hilbert series and taking the hyperK\"ahler quotient over the $U(1)^{n-1}$ topological symmetry. 

If instead, we were to gauge the entire diagonal $SU(n)$ in the flavor group, we can propose the following 3d mirror:
\begin{equation}
  \scalebox{0.8}{\begin{tikzpicture}
	\begin{pgfonlayer}{nodelayer}
		\node [style=gauge3] (0) at (-1.75, -3.75) {};
		\node [style=gauge3] (1) at (-0.75, -3.75) {};
		\node [style=gauge3] (2) at (1.25, -3.75) {};
		\node [style=flavour2] (3) at (2.25, -3.75) {};
		\node [style=none] (4) at (-0.25, -3.75) {};
		\node [style=none] (5) at (0.75, -3.75) {};
		\node [style=none] (6) at (0.25, -3.75) {$\dots$};
		\node [style=none] (7) at (-1.75, -4.25) {1};
		\node [style=none] (8) at (-0.75, -4.25) {2};
		\node [style=none] (9) at (1.25, -4.25) {$n-1$};
		\node [style=none] (10) at (2.25, -4.25) {$n$};
		\node [style=gauge3] (11) at (7.25, -3.75) {};
		\node [style=gauge3] (12) at (6.25, -3.75) {};
		\node [style=gauge3] (13) at (4.25, -3.75) {};
		\node [style=flavour2] (14) at (3.25, -3.75) {};
		\node [style=none] (15) at (5.75, -3.75) {};
		\node [style=none] (16) at (4.75, -3.75) {};
		\node [style=none] (17) at (5.25, -3.75) {$\dots$};
		\node [style=none] (18) at (7.25, -4.25) {1};
		\node [style=none] (19) at (6.25, -4.25) {2};
		\node [style=none] (20) at (4.25, -4.25) {$n-1$};
		\node [style=none] (21) at (3.25, -4.25) {$n$};
		\node [style=gauge3] (22) at (2.75, 1) {};
		\node [style=gauge3] (23) at (2.75, 0) {};
		\node [style=gauge3] (24) at (2.75, -2) {};
		\node [style=flavour2] (25) at (2.75, -3) {};
		\node [style=none] (26) at (2.75, -0.5) {};
		\node [style=none] (27) at (2.75, -1.5) {};
		\node [style=none] (28) at (2.75, -1) {$\vdots$};
		\node [style=none] (29) at (3.25, 1) {1};
		\node [style=none] (30) at (3.25, 0) {2};
		\node [style=none] (31) at (3.5, -2) {$n-1$};
		\node [style=none] (32) at (3.25, -3) {$n$};
		\node [style=none] (33) at (2.75, -2.5) {};
		\node [style=none] (34) at (1.75, -3.5) {};
		\node [style=none] (35) at (3.75, -3.5) {};
		\node [style=none] (36) at (2.75, -4.5) {};
		\node [style=none] (37) at (2.75, -5) {\textcolor{brown}{Gauge $SU(n)$ flavor symmetry}};
		\node [style=none] (38) at (7.5, -11) {};
		\node [style=none] (39) at (9, -11) {};
		\node [style=none] (41) at (8.25, -10.5) {3d mirror};
		\node [style=gauge3] (42) at (10, -9.25) {};
		\node [style=gauge3] (43) at (11.25, -9.25) {};
		\node [style=gauge3] (44) at (13.25, -9.25) {};
		\node [style=flavour2] (45) at (14.25, -9.25) {};
		\node [style=none] (46) at (11.75, -9.25) {};
		\node [style=none] (47) at (12.75, -9.25) {};
		\node [style=none] (48) at (12.25, -9.25) {$\dots$};
		\node [style=gauge3] (53) at (10, -10.75) {};
		\node [style=gauge3] (54) at (11.25, -10.75) {};
		\node [style=gauge3] (55) at (13.25, -10.75) {};
		\node [style=flavour2] (56) at (14.25, -10.75) {};
		\node [style=none] (57) at (11.75, -10.75) {};
		\node [style=none] (58) at (12.75, -10.75) {};
		\node [style=none] (59) at (12.25, -10.75) {$\dots$};
		\node [style=gauge3] (64) at (10, -12.5) {};
		\node [style=gauge3] (65) at (11.25, -12.5) {};
		\node [style=gauge3] (66) at (13.25, -12.5) {};
		\node [style=flavour2] (67) at (14.25, -12.5) {};
		\node [style=none] (68) at (11.75, -12.5) {};
		\node [style=none] (69) at (12.75, -12.5) {};
		\node [style=none] (70) at (12.25, -12.5) {$\dots$};
		\node [style=gauge3] (81) at (-1, -11) {};
		\node [style=gauge3] (82) at (0, -11) {};
		\node [style=gauge3] (83) at (2, -11) {};
		\node [style=none] (85) at (0.5, -11) {};
		\node [style=none] (86) at (1.5, -11) {};
		\node [style=none] (87) at (1, -11) {$\dots$};
		\node [style=none] (88) at (-1, -11.5) {1};
		\node [style=none] (89) at (0, -11.5) {2};
		\node [style=none] (90) at (2, -11.5) {$n-1$};
		\node [style=gauge3] (92) at (6.5, -11) {};
		\node [style=gauge3] (93) at (5.5, -11) {};
		\node [style=gauge3] (94) at (3.5, -11) {};
		\node [style=none] (96) at (5, -11) {};
		\node [style=none] (97) at (4, -11) {};
		\node [style=none] (98) at (4.5, -11) {$\dots$};
		\node [style=none] (99) at (6.5, -11.5) {1};
		\node [style=none] (100) at (5.5, -11.5) {2};
		\node [style=none] (101) at (3.5, -11.5) {$n-1$};
		\node [style=gauge3] (103) at (2.75, -7.25) {};
		\node [style=gauge3] (104) at (2.75, -8.25) {};
		\node [style=gauge3] (105) at (2.75, -10.25) {};
		\node [style=none] (107) at (2.75, -8.75) {};
		\node [style=none] (108) at (2.75, -9.75) {};
		\node [style=none] (109) at (2.75, -9.25) {$\vdots$};
		\node [style=none] (110) at (3.25, -7.25) {1};
		\node [style=none] (111) at (3.25, -8.25) {2};
		\node [style=none] (112) at (3.5, -10.25) {$n-1$};
		\node [style=none] (120) at (0.5, -3) {};
		\node [style=none] (132) at (2.625, -5.75) {};
		\node [style=none] (133) at (2.625, -6.25) {};
		\node [style=none] (134) at (2.875, -6.25) {};
		\node [style=none] (135) at (2.875, -5.75) {};
		\node [style=gauge3] (136) at (2.75, -11) {};
		\node [style=none] (137) at (2.75, -11.55) {$n$};
		\node [style=none] (138) at (10, -9.75) {\textcolor{red}{1}};
		\node [style=none] (139) at (11.25, -9.75) {\textcolor{red}{2}};
		\node [style=none] (140) at (13.25, -9.75) {\textcolor{red}{$n-1$}};
		\node [style=none] (141) at (10, -11.25) {\textcolor{red}{1}};
		\node [style=none] (142) at (11.25, -11.25) {\textcolor{red}{2}};
		\node [style=none] (143) at (13.25, -11.25) {\textcolor{red}{$n-1$}};
		\node [style=none] (144) at (10, -13) {\textcolor{red}{1}};
		\node [style=none] (145) at (11.25, -13) {\textcolor{red}{2}};
		\node [style=none] (146) at (13.25, -13) {\textcolor{red}{$n-1$}};
		\node [style=none] (147) at (14.25, -9.75) {$n$};
		\node [style=none] (148) at (14.25, -11.25) {$n$};
		\node [style=none] (149) at (14.25, -13) {$n$};
		\node [style=none] (150) at (15.25, -12.25) {};
		\node [style=none] (151) at (14, -13.75) {};
		\node [style=none] (152) at (15.5, -13.5) {\textcolor{red}{Gauge $SU(n)$ }};
		\node [style=none] (153) at (15, -14) {\textcolor{red}{ topological symmetry}};
	\end{pgfonlayer}
	\begin{pgfonlayer}{edgelayer}
		\draw (0) to (1);
		\draw (1) to (4.center);
		\draw (5.center) to (3);
		\draw (11) to (12);
		\draw (12) to (15.center);
		\draw (16.center) to (14);
		\draw (22) to (23);
		\draw (23) to (26.center);
		\draw (27.center) to (25);
		\draw [style=browndotted, bend right=45] (34.center) to (36.center);
		\draw [style=browndotted, bend right=45] (36.center) to (35.center);
		\draw [style=browndotted, bend right=45] (35.center) to (33.center);
		\draw [style=browndotted, bend left=315] (33.center) to (34.center);
		\draw [style=arrowed] (38.center) to (39.center);
		\draw [style=arrowed] (39.center) to (38.center);
		\draw (42) to (43);
		\draw (43) to (46.center);
		\draw (47.center) to (45);
		\draw (53) to (54);
		\draw (54) to (57.center);
		\draw (58.center) to (56);
		\draw (64) to (65);
		\draw (65) to (68.center);
		\draw (69.center) to (67);
		\draw (81) to (82);
		\draw (82) to (85.center);
		\draw (92) to (93);
		\draw (93) to (96.center);
		\draw (103) to (104);
		\draw (104) to (107.center);
		\draw (108.center) to (105);
		\draw (135.center) to (134.center);
		\draw (133.center) to (132.center);
		\draw (83) to (136);
		\draw (86.center) to (83);
		\draw (136) to (105);
		\draw (136) to (94);
		\draw [style=rede] (150.center) to (151.center);
	\end{pgfonlayer}
\end{tikzpicture}
}
\end{equation}
The theory on the right with the $SU(n)$ topological symmetry gauged is the well known $4d$ $\mathcal{N}=2$ class $\mathcal{S}$ theory called $T_n$ compactified on a circle. The 3d mirror relation to the three-legged non-linear quiver on the left is well known  \cite{Benini:2010uu}. To check this mirror relation, we perform an analogous computation as we did earlier in the section by first computing the refined Hilbert series (of the quiver on the right) and then taking the $SU(n)$ hyperK\"ahler quotient over the $SU(n)$ topological symmetry (as we did in \eqref{eq:Contraction3dMirror} for the $U(1)^{n-1}$). Due to the self-mirror nature of $T[SU(n)]$, it is trivial that the two computations will give the same answer. What could be more interesting is to gauge a subgroup $U(1)^{n-1} \subset H \subset SU(n)$ instead and finding its 3d mirror. If $H$ is a product of special unitary gauge groups (like the partial contraction case), one can try to find connections with class $\mathcal{S}$ theories when uplifted to 4d $\mathcal{N}=2$.

Another interesting question is at what point does the 3d mirror stop being a Lagrangian (quiver gauge theory). In the case where $H=U(1)^{n-1}$, one can still obtain the 3d mirror with $S[\prod_i \Un(n_i)]$ gauge groups as a Lagrangian/quiver theory. Whereas for $H=SU(n)$, the mirror theory is the non-Lagrangian $T_n$ theory on a circle.  This might hint at a fundamental limitation on what kind of 3d mirror pair a non-linear quiver can possess. In particular, whether the mirror is forbidden to be a quiver like theory with explicit gauge and flavor groups and one must resort to gauging topological symmetries.

\section{Ortho-Symplectic Contractions (D-type)}\label{sec:3}
Following the orthosymplectic implosion constructions in \cite{Bourget:2021zyc} and partial implosion in \cite{Bourget:2021qpx}, one can perform the same contractions for these quivers as well.

For simplicity we only focus on $G=SO(2n)$.\footnote{The magnetic quivers for $B$-type and $C$-type universal implosions were discussed along $D$-type (and $A$-type) in \cite{Bourget:2021zyc}. For $C$-type however the proposed quivers are \emph{bad} (in the classification of \cite{Gaiotto:2008ak}) and hence we cannot perform monopole formula computations. In \cite{Bourget:2021qpx} magnetic quivers were only proposed for $A$- and $D$-type partial implosions. We therefore leave the investigation of $B$- and $C$-type (partial) contractions for future work.}

\paragraph{Universal Implosion}
The magnetic quiver construction for the universal implosion $Q_G$ for $G=SO(2n)$ (with $T=U(1)^n$) is \cite{Bourget:2021zyc}:
\begin{equation}
    \uri=\mathcal{C}\left(\raisebox{-.5\height}{\begin{tikzpicture}
        \node[gauger,label=below:{\small$2$}] (-1) at (-1,0) {};
        \node[gaugeb,label=below:{\small$2$}] (0) at (0,0) {};
        \node[gauger,label=below:{\small$4$}] (1) at (1,0) {};
        \node[gaugeb,label=below:{\small$4$}] (2) at (2,0) {};
        \node (3) at (3,0) {$\cdots$};
        \node[gaugeb,rotate=-35,label=below:{\small$2n-2$}] (4) at (4,0) {};
        \draw (-1)--(0)--(1)--(2)--(3)--(4);
        \node[gauge,label=right:{\small$1$}] (5u) at (5,1.5) {};
        \node at (5,0.5) {$\vdots$};
        \node[gauge,label=right:{\small$1$}] (5m) at (5,-0.5) {};
        \draw (4)--(5u) (4)--(5m);
        \draw [decorate,decoration={brace,amplitude=5pt},xshift=0,yshift=0]
    (5.5,1.8)--(5.5,-0.8) node [black,midway,xshift=0.7cm,yshift=0] {$n$};
        \draw (6,-1)--(7,0);
        \node at (7,-1) {$\mathbb{Z}_2$};
    \end{tikzpicture}}\right)\;,
\end{equation}
where the colouring of the nodes is explained in Appendix \ref{app:Quivers}. The flavour symmetry of $Q_{G}$ is $F_{Q_{G}}=SO(2n)\times U(1)^n$.

\paragraph{Partial Implosion}
The magnetic quiver construction for the partial implosion $Q_{G,P}$ for $G=SO(2n)$ and $L_P=\prod_{i=1}^{l} GL_{k_i}\times SO(2k)$ (where $k\in\mathbb{N}\backslash\{1\}$\footnote{Note that since
\begin{equation}
    \raisebox{-0.6\height}{\begin{tikzpicture}
    \node[gauger,label=below:{$2$}] at (0,0) {};
\end{tikzpicture}}\quad\cong\quad\raisebox{-0.6\height}{\begin{tikzpicture}
    \node[gauge,label=below:{$1$}] at (0,0) {};
\end{tikzpicture}}\;,
\end{equation}
we don't consider the case of $k=1$.} and $(\sum_{i=1}^{l}2k_i)+2k=2n$) is \cite{Bourget:2021qpx}:
\begin{equation}
    \urip=\mathcal{C}\left(\raisebox{-.5\height}{\begin{tikzpicture}
        \node[gauger,label=below:{\small$2$}] (-1) at (-1,0) {};
        \node[gaugeb,label=below:{\small$2$}] (0) at (0,0) {};
        \node[gauger,label=below:{\small$4$}] (1) at (1,0) {};
        \node[gaugeb,label=below:{\small$4$}] (2) at (2,0) {};
        \node (3) at (3,0) {$\cdots$};
        \node[gaugeb,rotate=-10,label=below:{\small$2n-2$}] (4) at (4,0) {};
        \draw (-1)--(0)--(1)--(2)--(3)--(4);
        \node[gauge,label=above:{\small$k_1$}] (5u) at (5,1.5) {};
        \node at (5,0.5) {$\vdots$};
        \node[gauge,label=above:{\small$k_{1}-1$}] (6m) at (6,1.5) {};
        \node (7u) at (7,1.5) {$\cdots$};
        \node[gauge,label=above:{\small$2$}] (8u) at (8,1.5) {};
        \node[gauge,label=above:{\small$1$}] (9u) at (9,1.5) {};
        \node[gauge,label=above:{\small$k_{l}$}] (5m) at (5,-0.5) {};
        \node[gauge,label=above:{\small$k_{l}-1$}] (6m) at (6,-0.5) {};
        \node (7m) at (7,-0.5) {$\cdots$};
        \node[gauge,label=above:{\small$2$}] (8m) at (8,-0.5) {};
        \node[gauge,label=above:{\small$1$}] (9m) at (9,-0.5) {};
        \node[gauger,label=below:{\small$2k$}] (5d) at (5,-1.5) {};
        \node[gaugeb,label=below:{\small$2k-2$}] (6d) at (6,-1.5) {};
        \node (7d) at (7,-1.5) {$\cdots$};
        \node[gaugeb,label=below:{\small$2$}] (8d) at (8,-1.5) {};
        \node[gauger,label=below:{\small$2$}] (9d) at (9,-1.5) {};
        \draw (4)--(5u)--(6u)--(7u)--(8u)--(9u) (4)--(5m)--(6m)--(7m)--(8m)--(9m) (4)--(5d)--(6d)--(7d)--(8d)--(9d);
        \draw (9.5,-1)--(10.5,0);
        \node at (10.5,-1) {$\mathbb{Z}_2$};
    \end{tikzpicture}}\right)\;.
\end{equation}
The flavour symmetry of $Q_{G,P}$ is $F_{Q_{G,P}}=SO(2n)\times (\prod_{i=1}^l U(k_l))\times SO(2k)$.

\paragraph{Universal Contraction}
The universal contraction for $G=SO(2n)$,
\begin{equation}
    \ucon=\uli\circ \uri=(\uli\times \uri)///T\;,
\end{equation}
has magnetic quiver construction
\begin{equation}
    \ucon=\mathcal{C}\left(\raisebox{-.5\height}{\begin{tikzpicture}
        \node[gauger,label=below:{\small$2$}] (-1) at (-1,0) {};
        \node[gaugeb,label=below:{\small$2$}] (0) at (0,0) {};
        \node[gauger,label=below:{\small$4$}] (1) at (1,0) {};
        \node[gaugeb,label=below:{\small$4$}] (2) at (2,0) {};
        \node (3) at (3,0) {$\cdots$};
        \node[gaugeb,rotate=-35,label=below:{\small$2n-2$}] (4) at (4,0) {};
        \draw (-1)--(0)--(1)--(2)--(3)--(4);
        \node[gauge,label=above:{\small$1$}] (5u) at (5,1.5) {};
        \node at (5,0.5) {$\vdots$};
        \node[gauge,label=below:{\small$1$}] (5m) at (5,-0.5) {};
        \draw (4)--(5u) (4)--(5m);
        \node[gauger,label=below:{\small$2$}] (-1r) at (11,0) {};
        \node[gaugeb,label=below:{\small$2$}] (0r) at (10,0) {};
        \node[gauger,label=below:{\small$4$}] (1r) at (9,0) {};
        \node[gaugeb,label=below:{\small$4$}] (2r) at (8,0) {};
        \node (3r) at (7,0) {$\cdots$};
        \node[gaugeb,rotate=35,label=below:{\small$2n-2$}] (4r) at (6,0) {};
        \draw (-1r)--(0r)--(1r)--(2r)--(3r)--(4r);
        \draw (4r)--(5u) (4r)--(5m);
        \draw (11,-1)--(12,0);
        \node at (12,-1) {$\mathbb{Z}_2$};
    \end{tikzpicture}}\right)\;.
    \label{4.5}
\end{equation}
The flavour symmetry of $\ucon$ is $F_{\ucon}=SO(2n)\times U(1)^n\times SO(2n)$. Reduction of $\ucon$ by $T=U(1)^n$ gives
\begin{equation}
    \ucon///T=\mathcal{C}\left(\raisebox{-.5\height}{\begin{tikzpicture}
        \node[gauger,label=below:{\small$2$}] (-1) at (-1,0) {};
        \node[gaugeb,label=below:{\small$2$}] (0) at (0,0) {};
        \node[gauger,label=below:{\small$4$}] (1) at (1,0) {};
        \node[gaugeb,label=below:{\small$4$}] (2) at (2,0) {};
        \node (3) at (3,0) {$\cdots$};
        \node[gaugeb,rotate=-35,label=below:{\small$2n-2$}] (4) at (4,0) {};
        \draw (-1)--(0)--(1)--(2)--(3)--(4);
        \node[flavourr,label=below:{\small$2n$}] (5) at (5,0) {};
        \draw (4)--(5);
        \node[gauger,label=below:{\small$2$}] (-1r) at (11,0) {};
        \node[gaugeb,label=below:{\small$2$}] (0r) at (10,0) {};
        \node[gauger,label=below:{\small$4$}] (1r) at (9,0) {};
        \node[gaugeb,label=below:{\small$4$}] (2r) at (8,0) {};
        \node (3r) at (7,0) {$\cdots$};
        \node[gaugeb,rotate=35,label=below:{\small$2n-2$}] (4r) at (6,0) {};
        \draw (-1r)--(0r)--(1r)--(2r)--(3r)--(4r);
        \draw (4r)--(5);
    \end{tikzpicture}}\right)=\mathcal{N}^2\;,
\end{equation}
which is the product of two nilpotent cones $\mathcal{N}$ of $SO(2n)$, as expected.

One can show this by explicitly computing the Hilbert series. The first few orders of the Hilbert series of \eqref{4.5} for  $n=2$ is: 
\begin{equation}
    HS(z_1,z_2,z_3,t)=1+33t^2+559t^4+6157t^6+(47726+\frac{36}{z_1}+36z_1+\frac{36}{z_2}+36z_2+\frac{36}{z_3}+36z_3)t^4+O\left(t^{10}\right)
\end{equation}
where $z_1,z_2,z_3$ are the fugacities for the three $U(1)$s. Performing the hyperK\"ahler quotient over the three $U(1)$s gives:
\begin{equation}
    HS(t)=\oint \frac{dz_1dz_2dz_3}{(2\pi i)^3z_1z_2z_3}HS(z_1,z_2,z_3,t)=1+30t^2+463t^4+4578t^6+30899t^8 + O\left(t^{10}\right)
\end{equation}
which is the square of the Hilbert series of the nilcone of $SO(6)$. 

\paragraph{Partial Contraction}
The partial contraction for $G=SO(2n)$ and $L_P=\prod_{i=1}^{l} GL_{k_i}\times SO(2k)$,
\begin{equation}
    \uconp=\ulip\circ \urip=(\ulip\times \urip)///L_P\;,
\end{equation}
has magnetic quiver construction
\begin{equation}
    \uconp=\mathcal{C}\left(\raisebox{-.5\height}{\begin{tikzpicture}
        \node[gauger,label=below:{\small$2$}] (-1) at (-1,0) {};
        \node[gaugeb,label=below:{\small$2$}] (0) at (0,0) {};
        \node[gauger,label=below:{\small$4$}] (1) at (1,0) {};
        \node[gaugeb,label=below:{\small$4$}] (2) at (2,0) {};
        \node (3) at (3,0) {$\cdots$};
        \node[gaugeb,rotate=-35,label=below:{\small$2n-2$}] (4) at (4,0) {};
        \draw (-1)--(0)--(1)--(2)--(3)--(4);
        \node[gauge,label=above:{\small$k_1$}] (5u) at (5,1.5) {};
        \node at (5,0.5) {$\vdots$};
        \node[gauge,label=above:{\small$k_l$}] (5m) at (5,-0.5) {};
        \node[gauger,label=below:{\small$2k$}] (5d) at (5,-1.5) {};
        \draw (4)--(5u) (4)--(5m) (4)--(5d);
        \node[gauger,label=below:{\small$2$}] (-1r) at (11,0) {};
        \node[gaugeb,label=below:{\small$2$}] (0r) at (10,0) {};
        \node[gauger,label=below:{\small$4$}] (1r) at (9,0) {};
        \node[gaugeb,label=below:{\small$4$}] (2r) at (8,0) {};
        \node (3r) at (7,0) {$\cdots$};
        \node[gaugeb,rotate=35,label=below:{\small$2n-2$}] (4r) at (6,0) {};
        \draw (-1r)--(0r)--(1r)--(2r)--(3r)--(4r);
        \draw (4r)--(5u) (4r)--(5m) (4r)--(5d);
        \draw (11,-1)--(12,0);
        \node at (12,-1) {$\mathbb{Z}_2$};
    \end{tikzpicture}}\right)\;.
\end{equation}
The flavour symmetry of $\uconp$ is $F_{\uconp}=SO(2n)\times U(1)^l\times SO(2n)$. The reduction by $U(1)^l$ is
\begin{equation}
    \uconp=\mathcal{C}\left(\raisebox{-.5\height}{\begin{tikzpicture}
        \node[gauger,label=below:{\small$2$}] (-1) at (-1,0) {};
        \node[gaugeb,label=below:{\small$2$}] (0) at (0,0) {};
        \node[gauger,label=below:{\small$4$}] (1) at (1,0) {};
        \node[gaugeb,label=below:{\small$4$}] (2) at (2,0) {};
        \node (3) at (3,0) {$\cdots$};
        \node[gaugeb,rotate=-35,label=below:{\small$2n-2$}] (4) at (4,0) {};
        \draw (-1)--(0)--(1)--(2)--(3)--(4);
        \node[gauge,label=above:{\small$SU(k_1)$}] (5u) at (5,1.5) {};
        \node at (5,0.5) {$\vdots$};
        \node[gauge,label=above:{\small$SU(k_l)$}] (5m) at (5,-0.5) {};
        \node[gauger,label=below:{\small$2k$}] (5d) at (5,-1.5) {};
        \draw (4)--(5u) (4)--(5m) (4)--(5d);
        \node[gauger,label=below:{\small$2$}] (-1r) at (11,0) {};
        \node[gaugeb,label=below:{\small$2$}] (0r) at (10,0) {};
        \node[gauger,label=below:{\small$4$}] (1r) at (9,0) {};
        \node[gaugeb,label=below:{\small$4$}] (2r) at (8,0) {};
        \node (3r) at (7,0) {$\cdots$};
        \node[gaugeb,rotate=35,label=below:{\small$2n-2$}] (4r) at (6,0) {};
        \draw (-1r)--(0r)--(1r)--(2r)--(3r)--(4r);
        \draw (4r)--(5u) (4r)--(5m) (4r)--(5d);
        \draw (11,-1)--(12,0);
        \node at (12.4,-1) {$\mathbb{Z}_{\mathrm{gcd}(2,k_i)}$};
    \end{tikzpicture}}\right)\;.
\end{equation} 
 
\section*{Acknowledgements}

We would like to thank Antoine Bourget for discussions.  AD and JM would like to thank the Isaac Newton Institute for Mathematical Sciences, Cambridge, for support and hospitality during the programme \emph{New equivariant methods in algebraic and differential geometry} where work on this paper was undertaken. This work was supported by EPSRC grant no EP/R014604/1.
JFG is supported by the EPSRC Open Fellowship (Schafer-Nameki) EP/X01276X/1 and the ``Simons Collaboration on Special Holonomy in Geometry, Analysis and Physics''.  ZZ is supported by the ERC Consolidator Grant \# 864828 ``Algebraic Foundations of Supersymmetric Quantum Field Theory'' (SCFTAlg).

\clearpage

\appendix

\section{Unframed Quivers and Choice of Gauge Group}
\label{app:Quivers}

In this paper we deal mostly with unframed unitary or (special) unitary quivers, and some orthosymplectic quivers. We assume the reader is familiar with these concepts, however it is useful to recall some basic notions, see for example \cite{Bourget:2020xdz}.

\paragraph{Quiver Notation.}
We use square nodes for flavour nodes, round nodes for gauge nodes, white for unitary groups (unless specified to be special unitary), red for special orthogonal gauge groups and blue for compact symplectic groups.
\begin{equation}
    \begin{tikzpicture}
        \node[gauge,label=below:{$n$}] at (0,0) {};
        \node[gauge,label=below:{SU($n$)}] at (0,-1.5) {};
        \node[gauger,label=below:{$n$}] at (0,-3) {};
        \node[gaugeb,label=below:{$n$}] at (0,-4.5) {};
        \node at (1,0) {$=$};
        \node at (1,-1.5) {$=$};
        \node at (1,-3) {$=$};
        \node at (1,-4.5) {$=$};
        \node[gauge,label=below:{U($n$)}] at (2,0) {};
        \node[gauge,label=below:{SU($n$)}] at (2,-1.5) {};
        \node[gauger,label=below:{SO($n$)}] at (2,-3) {};
        \node[gaugeb,label=below:{USp($n$)}] at (2,-4.5) {};
    \end{tikzpicture}\;.
\end{equation}

\paragraph{Unitary Quivers.}
Take a (connected) unframed unitary quiver $\mathsf{Q}$ with $N$ nodes of ranks $(r_i)_{i=1,\dots,N}$. Call $\tilde{G}=\prod_i \Un(r_i)$ the naive gauge group. There is a $\Un(1)\subset\tilde{G}$ which acts trivially on the hypermultiplets. We take the actual gauge group of the quiver to be $G=\tilde{G}/\mathrm{U}(1)$, decoupling the trivially acting $\Un(1)$\footnote{In the maths literature this is sometimes referred to as the Crawley-Boevey trick referring to an observation in \cite{Crawley-Boevey2001}.}. One would naively associate to every $\Un(r_i)$ factor in $\tilde{G}$ a $\Un(1)$ factor in the torus of the Coulomb branch global symmetry. However, since we decoupled the overall $\Un(1)$, there are only $N-1$ $\Un(1)$ factors in the torus of the Coulomb branch global symmetry. Where helpful, we will explicitly denote the decoupling of the $\Un(1)$ as
\begin{equation}
    \begin{tikzpicture}
        \node at (0,0) {$\mathsf{Q}$};
        \draw (0,-0.7)--(1,0.3);
        \node at (1,-0.5) {$\Un(1)$.};
    \end{tikzpicture}\;.
\end{equation}
However in the main text we always assume the decoupling to take place, even if it is not explicitly stated.

Importantly, when we consider the union of two disjoint quivers, then each quiver has its own $\Un(1)$ in need of decoupling.

\paragraph{(Special) Unitary Quivers.}
Now consider a (connected) unframed quiver $\mathsf{Q}^{\mathrm{su}}$ with both $N$ unitary and $M>0$ special unitary nodes. Let the ranks of the special unitary nodes be $(r^{\mathrm{su}}_j-1)_{j=1,\dots,M}$. Again we may call $\tilde{G}=\prod_i \Un(r_i)\times\prod_j \SU(r^{\mathrm{su}}_j)$ the naive gauge group. Let $g=\mathrm{gcd}(r^{\mathrm{su}}_j)$. There is a $\mathbb{Z}_g\subset\tilde{G}$ which acts trivially on the hypermultiplets. We may \emph{pick} any divisor $k|g$ and specify the actual gauge group to be $G^k=\tilde{G}/\mathbb{Z}_k$.\footnote{$\mathbb{Z}_g$ is called a 1-form symmetry in the physics literature, and changing the gauge group from $\tilde{G}$ to $\tilde{G}/\mathbb{Z}_k$ is achieved by gauging the subgroup $\mathbb{Z}_k\subset\mathbb{Z}_g$ of the 1-form symmetry.} This does not affect the Higgs branch of the quiver, but it does affect the Coulomb branch. We denote the choice of $G^k$ by
\begin{equation}
    \begin{tikzpicture}
        \node at (0,0) {$\mathsf{Q}^{\mathrm{su}}$};
        \draw (0,-0.7)--(1,0.3);
        \node at (1,-0.5) {$\mathbb{Z}_k$.};
    \end{tikzpicture}\;
\end{equation}

\paragraph{Orthosymplectic Quivers.} For an unframed orthosymplectic quiver, there is also a choice of gauge group if the nodes involved are only USp, SO(even), and U. One may either take the naive gauge group $G$, or $G/\mathbb{Z}_2$. If there are special unitary nodes, and $g=\mathrm{gcd}(r^{\mathrm{su}}_j)$ is even, then there are the same two choices. If $g$ is odd, the only choice is $G$.

\section{Hyper-K\"ahler Quotients of Coulomb Branch Symmetries}
\label{app:HKQ}

Hyper-K\"ahler quotients of Coulomb branches are generally difficult to perform. Recently there has been a first step in systematic exploration \cite{Hanany:2023tvn}, which addresses the special case where the symmetry to be quotiented is realised on a `long leg' of a unitary quiver. Hyper-K\"ahler quotients by a diagonal symmetry of the product of two Coulomb branches with individual magnetic quivers, the case most relevant for our work, have been employed in the original work on 3d Mirrors of class-S theories \cite{Benini:2010uu}, and for example in \cite[Figure 5]{Bourget:2021jwo}.

We will show how to perform the hyper-K\"ahler quotients needed in the present paper.

\subsection{Hyper-K\"ahler Quotient by Diagonal Symmetry of two Legs}
\label{app:HKQ1}

In this section we will focus purely on the case of good or ugly unframed unitary quivers with legs of the kind
\begin{equation}
\label{eq:HKQ_leg}
    \begin{tikzpicture}
        \node (-1) at (-1,0) {$\cdots$};
        \node[gauge,label=below:{$n$}] (0) at (0,0) {};
        \node[gauge,label=below:{$n-1$}] (1) at (1,0) {};
        \node (2) at (2,0) {$\cdots$};
        \node[gauge,label=below:{$2$}] (3) at (3,0) {};
        \node[gauge,label=below:{$1$,}] (4) at (4,0) {};
        \draw (-1)--(0)--(1)--(2)--(3)--(4);
        \draw[thick,dashed,red] \convexpath{0,4}{0.25cm};
        \node at (2,0.5) {\color{red}$\un(1)\times\su(n)$};
    \end{tikzpicture}\;
\end{equation}
where the node of rank $n$ may have any balance above $-2$. There is a $\un(1)\times\su(n)$ subalgebra of the Coulomb branch global symmetry algebra of the quiver, where the $\un(1)$ part comes from the rank $n$ node, and the $\su(n)$ part comes from the tail of nodes of rank $n-1$ and less, which are all balanced.

We will only take hyper-K\"ahler quotients by the diagonal symmetry of two such legs, either connected to two disjoint quivers, or connected to the same quiver. Each deserves its own discussion.

\paragraph{Disjoint quivers.}

Take two good or ugly unframed unitary quivers, which both have a leg of the type \eqref{eq:HKQ_leg}.
\begin{equation}
\label{eq:HKQ_join_leg_1}
    \begin{tikzpicture}
        \node (-1) at (-1,0) {$\mathsf{Q}_1$};
        \node[gauge,label=below:{$n$}] (0) at (0,0) {};
        \node[gauge,label=below:{$n-1$}] (1) at (1,0) {};
        \node (2) at (2,0) {$\cdots$};
        \node[gauge,label=below:{$2$}] (3) at (3,0) {};
        \node[gauge,label=below:{$1$}] (4) at (4,0) {};
        \draw (-1)--(0)--(1)--(2)--(3)--(4);
        \node at (5,0) {$\oplus$};
        \node (11) at (11,0) {$\mathsf{Q}_2$};
        \node[gauge,label=below:{$n$}] (10) at (10,0) {};
        \node[gauge,label=below:{$n-1$}] (9) at (9,0) {};
        \node (8) at (8,0) {$\cdots$};
        \node[gauge,label=below:{$2$}] (7) at (7,0) {};
        \node[gauge,label=below:{$1$}] (6) at (6,0) {};
        \draw (11)--(10)--(9)--(8)--(7)--(6);
        \draw[thick,dashed,red] \convexpath{0,4}{0.25cm};
        \node at (2,0.5) {\color{red}$\un(1)_1\times\su(n)_1$};
        \draw[thick,dotted,red] \convexpath{6,10}{0.25cm};
        \node at (8,0.5) {\color{red}$\un(1)_2\times\su(n)_2$};
        \draw (-1,-0.7)--(-2,0.3);
        \node at (-2,-0.5) {$\Un(1)$};
        \draw (11,-0.7)--(12,0.3);
        \node at (12,-0.5) {$\Un(1)$.};
    \end{tikzpicture}\;
\end{equation}
We can perform a hyper-K\"ahler quotient by the diagonal subgroup $\SU(n)_{12}\subset$SU$(n)_1\times \SU(n)_2$, producing the quiver
\begin{equation}
\label{eq:HKQ_join_leg_2}
    \begin{tikzpicture}
        \node (-1) at (-1,0) {$\mathsf{Q}_1$};
        \node[gauge,label=below:{$n$}] (0) at (0,0) {};
        \node (1) at (1,0) {$\mathsf{Q}_2$};
        \draw (-1)--(0)--(1);
        \draw[thick,dashdotted,red] (0) circle (0.25);
        \node at (0,0.5) {\color{red}$\un(1)_{12}$};
        \draw (1,-0.7)--(2,0.3);
        \node at (2,-0.5) {$\Un(1)$.};
    \end{tikzpicture}\;
\end{equation}
Note that in \eqref{eq:HKQ_join_leg_1} we were dealing with two disconnected quivers, and therefore had to decouple two individual $\Un(1)$s. In \eqref{eq:HKQ_join_leg_2} however we are dealing with a single connected quiver, and hence have to decouple only a single $\Un(1)$. Going from \eqref{eq:HKQ_join_leg_1} to \eqref{eq:HKQ_join_leg_2} the quaternionic Coulomb branch dimension reduces by $n^2-1$, which is the dimension of $\SU(n)_{12}$.

\paragraph{Connected quiver.}

Take a connected good or ugly unframed quiver, with two legs of the type \eqref{eq:HKQ_leg}.
\begin{equation}
\label{eq:HKQ_join_leg_3}
    \begin{tikzpicture}
        \node[gauge,label=below:{$n$}] (0) at (0,0) {};
        \node[gauge,label=below:{$n-1$}] (1) at (1,0) {};
        \node (2) at (2,0) {$\cdots$};
        \node[gauge,label=below:{$2$}] (3) at (3,0) {};
        \node[gauge,label=below:{$1$}] (4) at (4,0) {};
        \draw (0)--(1)--(2)--(3)--(4);
        \node[gauge,label=below:{$n$}] (10) at (9,0) {};
        \node[gauge,label=below:{$n-1$}] (9) at (8,0) {};
        \node (8) at (7,0) {$\cdots$};
        \node[gauge,label=below:{$2$}] (7) at (6,0) {};
        \node[gauge,label=below:{$1$}] (6) at (5,0) {};
        \draw (10)--(9)--(8)--(7)--(6);
        \draw[thick,dashed,red] \convexpath{0,4}{0.25cm};
        \node at (2,0.5) {\color{red}$\un(1)_1\times\su(n)_1$};
        \draw[thick,dotted,red] \convexpath{6,10}{0.25cm};
        \node at (8,0.5) {\color{red}$\un(1)_2\times\su(n)_2$};
        \node (q) at (4.5,-2) {$\mathsf{Q}$};
        \draw (4.5,-3)--(5.5,-2.2);
        \node at (5.5,-2.8) {$\Un(1)$.};
        \draw (q) .. controls (-2,-2) and (-2,-1) .. (0);
        \draw (q) .. controls (11,-2) and (11,-1) .. (10);
    \end{tikzpicture}\;
\end{equation}
We can perform a hyper-K\"ahler quotient by the diagonal subgroup $\Un(n)_{12}\subset \Un(n)_1\times \Un(n)_2$, producing the quiver
\begin{equation}
\label{eq:HKQ_join_leg_4}
    \begin{tikzpicture}
        \node[gauge,label=below:{$n$}] (0) at (4.5,0) {};
        \draw[thick,dashdotted,red] (0) circle (0.25);
        \node at (4.5,0.5) {\color{red}$\un(1)_{21}$};
        \node (q) at (4.5,-2) {$\mathsf{Q}$};
        \draw (4.5,-3)--(5.5,-2.2);
        \node at (5.5,-2.8) {$\Un(1)$.};
        \draw (q) .. controls (1,-2) and (1,0) .. (0);
        \draw (q) .. controls (8,-2) and (8,0) .. (0);
    \end{tikzpicture}\;
\end{equation}
Going from \eqref{eq:HKQ_join_leg_3} to \eqref{eq:HKQ_join_leg_4} the quaternionic Coulomb branch dimension reduces by $n^2$, which is the dimension of $\Un(n)_{12}$. The $\un_{21}$ which remains as a Coulomb branch symmetry after the quotient is the antidiagonal $\Un(1)$ in $\Un(n)_1\times \Un(n)_2$ which survives the quotient by $\Un(n)_{12}$.

\subsection{Hyper-K\"ahler quotient of ${\Un(1)_{\mathrm{top}}}$ of a single node.}
\label{app:HKQ2}

In this section we will focus on good or ugly unframed (special) unitary quivers, and consider gauging the U$(1)_\mathrm{top}$ Coulomb branch symmetry associated to a unitary node, which is visible in the UV.

\paragraph{Unitary Quiver.}

First let's start with a purely unitary quiver with a node of rank $n$
\begin{equation}
    \begin{tikzpicture}
        \node[gauge,label=below:{$n$}] (0) at (4.5,0) {};
        \draw[thick,dashdotted,red] (0) circle (0.25);
        \node at (4.5,0.5) {\color{red}$\un(1)$};
        \node (q) at (4.5,-2) {$\mathsf{Q}$};
        \draw (4.5,-3)--(5.5,-2.2);
        \node at (5.5,-2.8) {$\Un(1)$,};
        \draw (q) .. controls (1,-2) and (1,0) .. (0);
        \draw (q) .. controls (3,-1.5) and (3,-0.5) .. (0);
        \node at (4.5,-1) {$\cdots$};
        \draw (q) .. controls (6,-1.5) and (6,-0.5) .. (0);
        \draw (q) .. controls (8,-2) and (8,0) .. (0);
    \end{tikzpicture}\;
\end{equation}
where we associate a $\un(1)$ Coulomb branch symmetry to the node in question, keeping in mind that one decouples an overall $\Un(1)$.

We can turn this quiver in an unframed (special) unitary quiver
\begin{equation}
    \begin{tikzpicture}
        \node[gauge,label=below:{$\SU(n)$}] (0) at (4.5,0) {};
        \node (q) at (4.5,-2) {$\mathsf{Q}$};
        \draw (4.5,-3)--(5.5,-2.2);
        \node at (5.5,-2.8) {$\mathbb{Z}_n$,};
        \draw (q) .. controls (1,-2) and (1,0) .. (0);
        \draw (q) .. controls (3,-1.5) and (3,-0.5) .. (0);
        \node at (4.5,-1) {$\cdots$};
        \draw (q) .. controls (6,-1.5) and (6,-0.5) .. (0);
        \draw (q) .. controls (8,-2) and (8,0) .. (0);
    \end{tikzpicture}\;
\end{equation}
which has the same Higgs and Coulomb branch. We refer to this operation as a hyper-K\"ahler quotient by $\SU(1)$, which is of course not meaningful as a geometric operation, but it is still a non-trivial operation on the quiver.

\paragraph{(Special) Unitary Quiver.}

Now take a (special) unitary quiver with $g=\mathrm{gcd}(r^{\mathrm{su}}_j)$, and a unitary node of rank $n$
\begin{equation}
    \begin{tikzpicture}
        \node[gauge,label=below:{$n$}] (0) at (4.5,0) {};
        \draw[thick,dashdotted,red] (0) circle (0.25);
        \node at (4.5,0.5) {\color{red}$\un(1)$};
        \node (q) at (4.5,-2) {$\mathsf{Q}^{\mathrm{su}}$};
        \draw (4.5,-3)--(5.5,-2.2);
        \node at (5.5,-2.8) {$\mathbb{Z}_g$,};
        \draw (q) .. controls (1,-2) and (1,0) .. (0);
        \draw (q) .. controls (3,-1.5) and (3,-0.5) .. (0);
        \node at (4.5,-1) {$\cdots$};
        \draw (q) .. controls (6,-1.5) and (6,-0.5) .. (0);
        \draw (q) .. controls (8,-2) and (8,0) .. (0);
    \end{tikzpicture}\;
\end{equation}
where the node in question carries a $\un(1)$ Coulomb branch symmetry, and there is no decoupling of an overall U$(1)$.

We can perform a hyper-K\"ahler quotient by U$(1)$, resulting in the (special) unitary quiver
\begin{equation}
    \begin{tikzpicture}
        \node[gauge,label=below:{$\SU(n)$}] (0) at (4.5,0) {};
        \node (q) at (4.5,-2) {$\mathsf{Q}^{\mathrm{su}}$};
        \draw (4.5,-3)--(5.5,-2.2);
        \node at (5.8,-2.8) {$\mathbb{Z}_{\mathrm{gcd}(g,n)}$.};
        \draw (q) .. controls (1,-2) and (1,0) .. (0);
        \draw (q) .. controls (3,-1.5) and (3,-0.5) .. (0);
        \node at (4.5,-1) {$\cdots$};
        \draw (q) .. controls (6,-1.5) and (6,-0.5) .. (0);
        \draw (q) .. controls (8,-2) and (8,0) .. (0);
    \end{tikzpicture}\;
\end{equation}
The quaternionic dimension of the Coulomb branch is decreased by 1 as expected. Furthermore, the centre of the gauge group remains unchanged, due to the quotient by $\mathbb{Z}_{\mathrm{gcd}(g,n)}$ in the latter quiver.

\subsection{Illustrative Examples}

In Figure \ref{fig:HKQ1} we give the simplest example relevant our paper of the hyper-K\"ahler quotients mentioned in Appendix \ref{app:HKQ1}. In Figure \ref{fig:HKQ2} we give the simplest example relevant to our paper of the hyper-K\"ahler quotients mentioned in Appendix \ref{app:HKQ2}.

\begin{landscape}

\begin{figure}[h]
    \centering
    \scalebox{0.7}{\begin{tikzpicture}
        \node (a) at (0,0) {$\begin{tikzpicture}
            \node[gauge,label=below:{1}] (1) at (1,0) {};
            \node[gauge,label=below:{2}] (2) at (2,0) {};
            \node[gauge,label=below:{2}] (3d) at (3,-1) {};
            \node[gauge,label=below:{1}] (4d) at (4,-1) {};
            \node[gauge,label=above:{1}] (3u) at (3,1) {};
            \draw (1)--(2)--(3d)--(4d) (2)--(3u);
            \node[gauge,label=below:{1}] (r1) at (9,0) {};
            \node[gauge,label=below:{2}] (r2) at (8,0) {};
            \node[gauge,label=below:{2}] (r3d) at (7,-1) {};
            \node[gauge,label=below:{1}] (r4d) at (6,-1) {};
            \node[gauge,label=above:{1}] (r3u) at (7,1) {};
            \draw (r1)--(r2)--(r3d)--(r4d) (r2)--(r3u);
            \draw (0.5,-0.5)--(1.5,-1.5);
            \node at (0.6,-1.4) {$\Un(1)$};
            \draw (9.5,-0.5)--(8.5,-1.5);
            \node at (9.4,-1.4) {$\Un(1)$};
            \node at (5,0) {$\oplus$};
            \draw[thick,dashed,blue] (3u) circle (0.25);
            \draw[thick,dotted,blue] (r3u) circle (0.25);
            \draw[thick,dashed,red] \convexpath{3d,4d}{0.25cm};
            \draw[thick,dotted,red] \convexpath{r3d,r4d}{0.25cm};
        \end{tikzpicture}$};
        \node (ams) at (8,2) {$\mathbb{H}^6\oplus\mathbb{H}^6$};
        \node (b) at (-8,-6) {$\begin{tikzpicture}
            \node[gauge,label=below:{1}] (1) at (1,0) {};
            \node[gauge,label=below:{2}] (2) at (2,0) {};
            \node[gauge,label=below:{2}] (3d) at (3,-1) {};
            \node[gauge,label=below:{1}] (4d) at (4,-1) {};
            \node[gauge,label=above:{1}] (3u) at (5,1) {};
            \draw (1)--(2)--(3d)--(4d) (2)--(3u);
            \node[gauge,label=below:{1}] (r1) at (9,0) {};
            \node[gauge,label=below:{2}] (r2) at (8,0) {};
            \node[gauge,label=below:{2}] (r3d) at (7,-1) {};
            \node[gauge,label=below:{1}] (r4d) at (6,-1) {};
            \draw (r1)--(r2)--(r3d)--(r4d) (r2)--(3u);
            \draw (9.5,-0.5)--(8.5,-1.5);
            \node at (9.4,-1.4) {$\Un(1)$};
            \draw[thick,dashed,red] \convexpath{3d,4d}{0.25cm};
            \draw[thick,dotted,red] \convexpath{r3d,r4d}{0.25cm};
        \end{tikzpicture}$};
        \node (c) at (8,-6) {$\begin{tikzpicture}
            \node[gauge,label=below:{1}] (1) at (1,0) {};
            \node[gauge,label=below:{2}] (2) at (2,0) {};
            \node[gauge,label=below:{2}] (3d) at (5,-1) {};
            \node[gauge,label=above:{1}] (3u) at (3,1) {};
            \draw (1)--(2)--(3d) (2)--(3u);
            \node[gauge,label=below:{1}] (r1) at (9,0) {};
            \node[gauge,label=below:{2}] (r2) at (8,0) {};
            \node[gauge,label=above:{1}] (r3u) at (7,1) {};
            \draw (r1)--(r2)--(3d) (r2)--(r3u);
            \draw (9.5,-0.5)--(8.5,-1.5);
            \node at (9.4,-1.4) {$\Un(1)$};
            \draw[thick,dashed,blue] (3u) circle (0.25);
            \draw[thick,dotted,blue] (r3u) circle (0.25);
        \end{tikzpicture}$};
        \node (cms) at (16,-4) {$\begin{tikzpicture}
            \node[gauge,label=below:{$\SU(2)$}] (0) at (0,0) {};
            \node[flavour,label=above:{$D_6$}] (1) at (0,1) {};
            \draw (0)--(1);
        \end{tikzpicture}$};
        \node (d) at (0,-12) {$\begin{tikzpicture}
            \node[gauge,label=below:{1}] (1) at (1,0) {};
            \node[gauge,label=below:{2}] (2) at (2,0) {};
            \node[gauge,label=below:{2}] (3d) at (5,-1) {};
            \node[gauge,label=above:{1}] (3u) at (5,1) {};
            \draw (1)--(2)--(3d) (2)--(3u);
            \node[gauge,label=below:{1}] (r1) at (9,0) {};
            \node[gauge,label=below:{2}] (r2) at (8,0) {};
            \draw (r1)--(r2)--(3d) (r2)--(3u);
            \draw (9.5,-0.5)--(8.5,-1.5);
            \node at (9.4,-1.4) {$\Un(1)$};
        \end{tikzpicture}$};
        \node (dms) at (9,-12) {$\begin{tikzpicture}
            \node[gauge,label=below:{$\Un(2)$}] (0) at (0,0) {};
            \node[flavour,label=above:{$6$}] (1) at (0,1) {};
            \draw (0)--(1);
        \end{tikzpicture}$};
        \draw[->] (a)--(b);
        \node at (-5.2,-3) {$/\!/\!/$S({\color{blue}$\Un(1)$})};
        \draw[->] (a)--(c);
        \node at (5.2,-3) {$/\!/\!/$S({\color{red}$\Un(2)$})};
        \draw[->] (b)--(d);
        \node at (-5.2,-9) {$/\!/\!/${\color{red}$\Un(2)$}};
        \draw[->] (c)--(d);
        \node at (5.2,-9) {$/\!/\!/${\color{blue}$\Un(1)$}};
        \draw[->] (a) .. controls (-1,-6) .. (d);
        \node at (0.5,-6) {$/\!/\!/$S({\color{blue}$\Un(1)$}{\color{red}$\Un(2)$})};
        \draw[goodgreen,<->] (a)--(ams);
        \node at (5.7,1.8) {\color{goodgreen}3d MS};
        \draw[goodgreen,<->] (c)--(cms);
        \node at (13.7,1.8-6) {\color{goodgreen}3d MS};
        \draw[goodgreen,<->] (d)--(dms);
        \node at (6.5,-11.5) {\color{goodgreen}3d MS};
        \draw[goodgreen,<->] (b) .. controls (-8,4) .. (ams);
    \end{tikzpicture}}
    \caption{Hyper-K\"ahler quotient of product of partial implosions by the diagonal of the Levy to reach the partial contraction, on the level of magnetic quivers and their 3d mirrors.}
    \label{fig:HKQ1}
\end{figure}
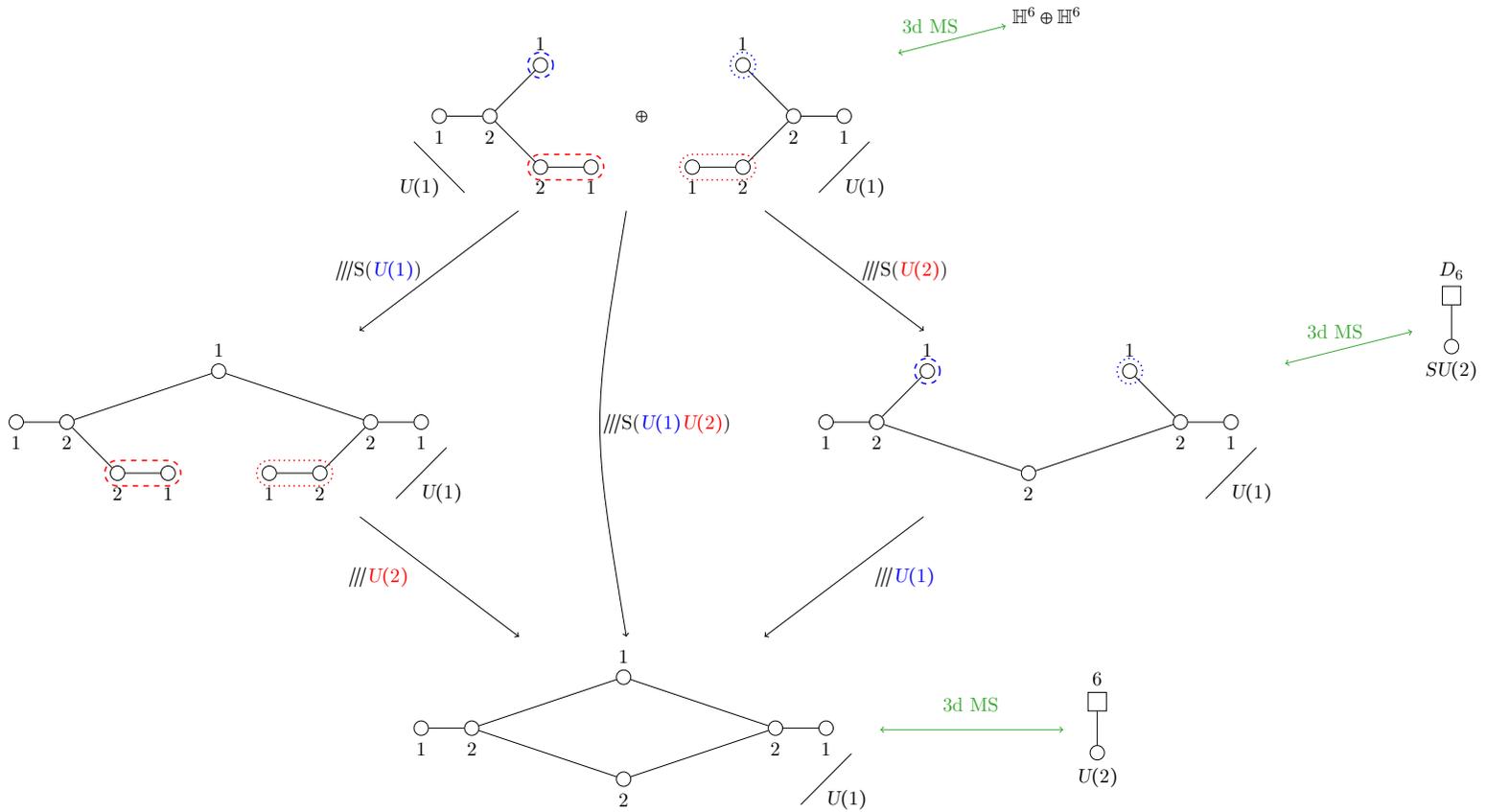

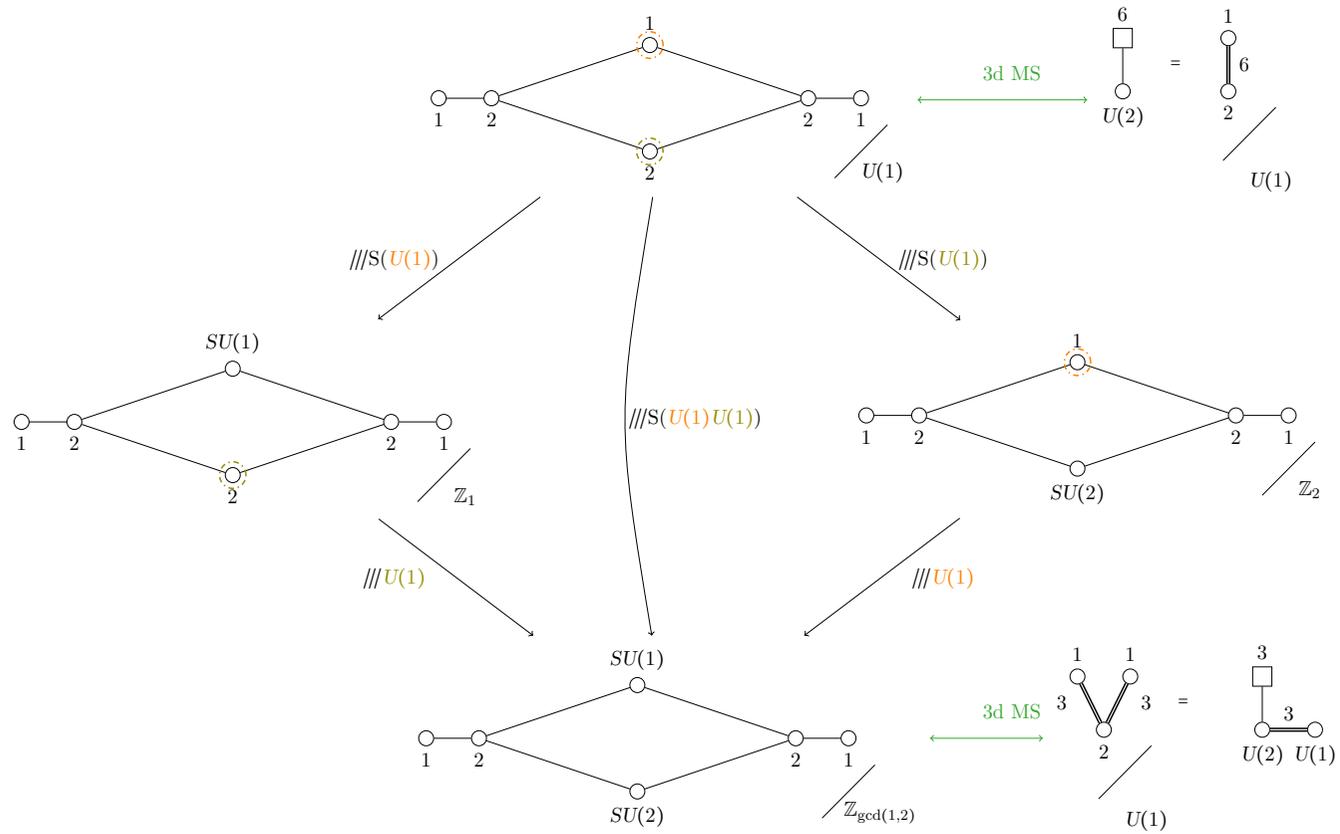
\begin{figure}[h]
    \centering
    \scalebox{0.7}{\begin{tikzpicture}
        \node (a) at (0,0) {$\begin{tikzpicture}
            \node[gauge,label=below:{1}] (1) at (1,0) {};
            \node[gauge,label=below:{2}] (2) at (2,0) {};
            \node[gauge,label=below:{2}] (3d) at (5,-1) {};
            \node[gauge,label=above:{1}] (3u) at (5,1) {};
            \draw (1)--(2)--(3d) (2)--(3u);
            \node[gauge,label=below:{1}] (r1) at (9,0) {};
            \node[gauge,label=below:{2}] (r2) at (8,0) {};
            \draw (r1)--(r2)--(3d) (r2)--(3u);
            \draw (9.5,-0.5)--(8.5,-1.5);
            \node at (9.4,-1.4) {$\Un(1)$};
            \draw[thick,dashdotted,orange] (3u) circle (0.25);
            \draw[thick,dashdotted,olive] (3d) circle (0.25);
        \end{tikzpicture}$};
        \node (ams) at (10,0) {$\begin{tikzpicture}
            \node[gauge,label=below:{$\Un(2)$}] (0) at (0,0) {};
            \node[flavour,label=above:{$6$}] (1) at (0,1) {};
            \draw (0)--(1);
            \node at (1,0.5) {$=$};
            \node[gauge,label=below:{2}] (0r) at (2,0) {};
            \node[gauge,label=above:{1}] (1r) at (2,1) {};
            \draw[thick,double] (0r)--(1r);
            \node at (2.3,0.5) {$6$};
            \draw (2.9,-0.3)--(1.9,-1.3);
            \node at (2.8,-1.7) {$\Un(1)$};
        \end{tikzpicture}$};
        \node (b) at (-8,-6) {$\begin{tikzpicture}
            \node[gauge,label=below:{1}] (1) at (1,0) {};
            \node[gauge,label=below:{2}] (2) at (2,0) {};
            \node[gauge,label=below:{2}] (3d) at (5,-1) {};
            \node[gauge,label=above:{$\SU(1)$}] (3u) at (5,1) {};
            \draw (1)--(2)--(3d) (2)--(3u);
            \node[gauge,label=below:{1}] (r1) at (9,0) {};
            \node[gauge,label=below:{2}] (r2) at (8,0) {};
            \draw (r1)--(r2)--(3d) (r2)--(3u);
            \draw (9.5,-0.5)--(8.5,-1.5);
            \node at (9.4,-1.4) {$\mathbb{Z}_{1}$};
            \draw[thick,dashdotted,olive] (3d) circle (0.25);
        \end{tikzpicture}$};
        \node (c) at (8,-6) {$\begin{tikzpicture}
            \node[gauge,label=below:{1}] (1) at (1,0) {};
            \node[gauge,label=below:{2}] (2) at (2,0) {};
            \node[gauge,label=below:{$\SU(2)$}] (3d) at (5,-1) {};
            \node[gauge,label=above:{1}] (3u) at (5,1) {};
            \draw (1)--(2)--(3d) (2)--(3u);
            \node[gauge,label=below:{1}] (r1) at (9,0) {};
            \node[gauge,label=below:{2}] (r2) at (8,0) {};
            \draw (r1)--(r2)--(3d) (r2)--(3u);
            \draw (9.5,-0.5)--(8.5,-1.5);
            \node at (9.4,-1.4) {$\mathbb{Z}_{2}$};
            \draw[thick,dashdotted,orange] (3u) circle (0.25);
        \end{tikzpicture}$};
        \node (d) at (0,-12) {$\begin{tikzpicture}
            \node[gauge,label=below:{1}] (1) at (1,0) {};
            \node[gauge,label=below:{2}] (2) at (2,0) {};
            \node[gauge,label=below:{$\SU(2)$}] (3d) at (5,-1) {};
            \node[gauge,label=above:{$\SU(1)$}] (3u) at (5,1) {};
            \draw (1)--(2)--(3d) (2)--(3u);
            \node[gauge,label=below:{1}] (r1) at (9,0) {};
            \node[gauge,label=below:{2}] (r2) at (8,0) {};
            \draw (r1)--(r2)--(3d) (r2)--(3u);
            \draw (9.5,-0.5)--(8.5,-1.5);
            \node at (9.6,-1.4) {$\mathbb{Z}_{\mathrm{gcd}(1,2)}$};
        \end{tikzpicture}$};
        \node (dms) at (10,-12) {$\begin{tikzpicture}
            \node[gauge,label=below:{2}] (0r) at (0,0) {};
            \node[gauge,label=above:{1}] (1rl) at (-0.5,1) {};
            \node[gauge,label=above:{1}] (1rr) at (0.5,1) {};
            \draw[thick,double] (1rl)--(0r)--(1rr);
            \node at (-0.8,0.5) {$3$};
            \node at (0.8,0.5) {$3$};
            \draw (0.9,-0.3)--(-0.1,-1.3);
            \node at (0.8,-1.7) {$\Un(1)$};
            \node at (1.5,0.5) {$=$};
            \node[gauge,label=below:{$\Un(2)$}] (0) at (3,0) {};
            \node[gauge,label=below:{$\Un(1)$}] (00) at (4,0) {};
            \node[flavour,label=above:{$3$}] (1) at (3,1) {};
            \draw (0)--(1);
            \draw[thick,double] (0)--(00);
            \node at (3.5,0.3) {3};
        \end{tikzpicture}$};
        \draw[->] (a)--(b);
        \node at (-5.2,-3) {$/\!/\!/$S({\color{orange}$\Un(1)$})};
        \draw[->] (a)--(c);
        \node at (5.2,-3) {$/\!/\!/$S({\color{olive}$\Un(1)$})};
        \draw[->] (b)--(d);
        \node at (-5.2,-9) {$/\!/\!/${\color{olive}$\Un(1)$}};
        \draw[->] (c)--(d);
        \node at (5.2,-9) {$/\!/\!/${\color{orange}$\Un(1)$}};
        \draw[->] (a) .. controls (-1,-6) .. (d);
        \node at (0.5,-6) {$/\!/\!/$S({\color{orange}$\Un(1)$}{\color{olive}$\Un(1)$})};
        \draw[goodgreen,<->] (a)--(ams);
        \node at (6.5,0.5) {\color{goodgreen}3d MS};
        \draw[goodgreen,<->] (d)--(dms);
        \node at (6.5,-11.5) {\color{goodgreen}3d MS};
        \draw[goodgreen,<->] (b) .. controls (-6,5) .. (ams);
        \draw[goodgreen,<->] (c) .. controls (10,-3) .. (ams);
        \node at (11,-3) {\color{goodgreen}3d MS};
    \end{tikzpicture}}
    \caption{Hyper-K\"ahler quotient or partial contraction by the center of the Levy, on the level of magnetic quivers and their 3d mirrors.}
    \label{fig:HKQ2}
\end{figure}

\end{landscape}

\bibliographystyle{JHEP}
\bibliography{bibli.bib}

\end{document}